%% file: main.tex
\documentclass[letterpaper]{IEEEtran}
\usepackage{cite}
\usepackage{babel}
\usepackage{amsmath,amssymb,amsfonts,amsthm}
\usepackage{siunitx}
\usepackage{graphicx}
\usepackage{textcomp}
\usepackage[capitalise]{cleveref}
\usepackage{color}
\usepackage{algorithm,algpseudocode}
\usepackage{tikz}
\usepackage{textcomp}

\include{macros}

\title{Multi-Group Multicast Beamforming by Superiorized Projections onto Convex Sets}
\author{Jochen~Fink,~\IEEEmembership{Student~Member,~IEEE,}
Renato~L.G.~Cavalcante,~\IEEEmembership{Member,~IEEE,}
and~S\l{}awomir Sta\'nczak,~\IEEEmembership{Senior~Member,~IEEE}% <-this % stops a space
\\
{\textit{Technische Universit\"at Berlin} and \textit{Fraunhofer Heinrich-Hertz-Institute}, Berlin, Germany}% <-this % stops a space
\\
{\{jochen.fink, renato.cavalcante, slawomir.stanczak\}@hhi.fraunhofer.de}% <-this % stops a space
\thanks{This work is supported by the Federal Ministry of Education and Research of the Federal Republic of Germany (BMBF) in the framework of the project AI4Mobile with funding number 16KIS1170K, and by the Federal Ministry for Economic Affairs and Energy of the Federal Republic of Germany (BMWi) in the framework of the project LIPS with funding number 01MD18010E.
The authors alone are responsible for the content of the paper.}}

\newcommand\copyrighttext{%
  \footnotesize
  This manuscript has been submitted to IEEE Transactions on Signal Processing for possible publication.
}
\newcommand\copyrightnotice{%
\begin{tikzpicture}[remember picture,overlay]
\node[anchor=south,yshift=10pt] at (current page.south) {\fbox{\parbox{\dimexpr\textwidth-\fboxsep-\fboxrule\relax}{\copyrighttext}}};
\end{tikzpicture}%
}

\begin{document}

\maketitle
\copyrightnotice{}
\input{00-abstract}
\input{01-introduction}
\input{02-problem_statement}

\input{03-algorithmic_solution}
\input{04-numerical_results}
\input{05-conclusion}
\input{06-appendix}

\bibliographystyle{IEEEbib}
\bibliography{refs}	
\end{document}

%% file: macros.tex
\def\a{{\mathbf a}}
\def\d{\mathbf{d}}

\def\u{{\mathbf u}}
\def\e{{\mathbf e}}
\def\v{{\mathbf v}}
\def\x{{\mathbf x}}
\def\y{{\mathbf y}}
\def\z{{\mathbf z}}
\def\h{{\mathbf h}}
\def\w{{\mathbf w}}

\def\Null{{\mathbf 0}}

\def\A{{\mathbf A}}
\def\B{{\mathbf B}}
\def\D{{\mathbf D}}
\def\E{\e_i\e^T}
\def\X{{\mathbf X}}
\def\U{{\mathbf U}}
\def\V{{\mathbf V}}
\def\W{{\mathbf W}}
\def\S{{\mathbf S}}

\def\SIG{{\mathbf \Sigma}}

\def\Y{{\mathbf Y}}
\def\Z{{\mathbf Z}}
\def\Q{{\mathbf Q}}
\def\I{{\mathbf I}}
\def\J{{\mathbf J}}

\def\P{{\mathbf P}}
\def\Z{{\mathbf Z}}
\def\EW{{\mathbf \Lambda}}
\def\EV{{\mathbf V}}

\def\CN{\mathcal{CN}}
\def\Group{\mathcal{G}}

\def\hilbert{{\mathcal H}}
\def\Hilbert{{{\mathcal H}^M}}
\def\RR{{\mathbb R}}
\def\CC{{\mathbb C}}
\def\NN{{\mathbb N}}

\def\trace{\mathrm{tr}}
\def\minimize{\mathrm{minimize}}
\def\argmin{\mathrm{arg\ min}}

\def\st{\mathrm{s.t.}}
\def\rank{\mathrm{rank}}
\def\diag{\mathrm{diag}}

\def\prox{{\mathrm{prox}}}

\def\Id{{\mathrm{Id}}}

\def\pPsd{P_{\hilbert_+}}

\def\muPsd{{\mu_{K+2}}}
\def\muP{{\mu_{K+1}}}
\def\muK{{\mu_K}}
\def\muOne{{\mu_1}}

\def\T[#1]{T_{#1}}
\def\P[#1]{P_{#1}}

\def\Xk[#1]{{\X_{#1}}}
\def\Yk[#1]{{\Y_{#1}}}

\def\Tstar{\T[\star]}
\def\Tp{\T[\mathrm{P}]}
\def\Tr{\P[\setR]}

\def\Re{\mathrm{Re}}
\def\Im{\mathrm{Im}}

\def\ipspacing{\hspace{-.15em}} % controls the spacing between double inner product signs
\def\Langle{\langle\ipspacing\langle}
\def\Rangle{\rangle\ipspacing\rangle}

\def\normbarspacing{-0.25ex} % controls the spacing between the 3 norm bars
\newcommand{\vertiii}[1]{{\left\vert\kern\normbarspacing\left\vert\kern\normbarspacing\left\vert #1 
		\right\vert\kern\normbarspacing\right\vert\kern\normbarspacing\right\vert}}
\def\Norm[#1]{\vertiii{#1}}
\def\current[#1]{{#1}^{(n)}}
\def\nextit[#1]{{#1}^{(n+1)}}
\def\zeroth[#1]{{#1}^{(0)}}

\def\relu[#1]{\left(#1\right)_+}

\def\fix{\mathrm{Fix}}

\def\setC{\mathcal{C}}
\def\setD{\mathcal{D}}
\def\setH{\mathcal{H}}
\def\setK{\mathcal{K}}
\def\setM{\mathcal{M}}
\def\setN{\mathcal{N}}
\def\setP{\mathcal{P}}
\def\setQ{\mathcal{Q}}
\def\setR{\mathcal{R}}

\def\setV{\mathcal{V}}
\def\setY{\mathcal{Y}}

\def\Cp{\setP}
\def\Cpsd{{\setC_+}}
\def\Cqk{{\setQ_k}}
\def\Cqk{{\setQ_k}}

\def\CqOne{{\setQ_1}}
\def\CqK{{\setQ_K}}

\def\supObj{\phi} % superiorization objective function (general)
\def\sobjf{f_1} % superiorization objective function (power reduction)
\def\sobjg{f_2} % superiorization objective function (rank reduction)

\def\bfscale{\rho(\w)} % scaling factor of beamforming vectors
\def\bound{R}
\def\nmax{n_{\max}}

\def\randA{\texttt{RandA}}
\def\sdrgau{\texttt{SDR-GauRan}}

\def\fppsca{\texttt{FPP-SCA}}
\def\admm{\texttt{CCP-ADMM}}
\def\proposed{\texttt{S-POCS}}
\def\SINRmin{\mathrm{SINR}^{\min}_{\rho}}
\def\Psdr{P_{\mathrm{SDR}}^\star}

\newtheorem{prop}{Proposition}
\newtheorem{remark}{Remark}
\newtheorem{lemma}{Lemma}
\newtheorem{fact}{Fact}
\newtheorem{example}{Example}
\newtheorem{definition}{Definition}

\renewcommand{\triangleq}{:=}
\def\plotScale{0.59}

%% file: 00-abstract.tex
\begin{abstract}
In this paper, we propose an iterative algorithm to address the nonconvex multi-group multicast beamforming problem with quality-of-service constraints and per-antenna power constraints. 
We formulate a convex relaxation of the problem as a semidefinite program in a real Hilbert space, which allows us to approximate a point in the feasible set by iteratively applying a bounded perturbation resilient fixed-point mapping. Inspired by the superiorization methodology, we use this mapping as a basic algorithm, and we add in each iteration a small perturbation with the intent to reduce the objective value and the distance to nonconvex rank-constraint sets.
We prove that the sequence of perturbations is bounded, so the algorithm is guaranteed to converge to a feasible point of the relaxed semidefinite program.
Simulations show that the proposed approach outperforms existing algorithms in terms of both computation time and approximation gap in many cases.
\end{abstract}

%% file: 01-introduction.tex
\section{Introduction}
\IEEEPARstart{M}{any} applications in wireless networks involve multicast communication, which can be defined as the transmission of identical information to multiple receivers. 
One example is connected driving, where applications such as platooning can benefit from transmitting the same status or control information to a group of vehicles \cite{zheng2015stability}.
Another example is the transmission of audio signals for live events, where each spectator can select from a variety of audio streams. 
Both use cases can benefit considerably from physical layer precoders that ensure a given quality-of-service (QoS) level for the requested stream at each receiver while reusing the same time and frequency resources for all receivers.

Physical layer multicasting schemes have been extensively investigated in the last two decades. The authors of \cite{sidiropoulos2006transmit} show that the performance of multicast transmission can be greatly improved by exploiting channel state information (CSI) at the transmitter. They consider two beamforming problems for single-group multicast beamforming, the max-min-fair (MMF) multicast beamforming problem and the QoS constrained multicast beamforming problem. While the MMF formulation aims at maximizing the lowest signal-to-noise ratio (SNR) among a group of users subject to a unit power constraint on the beamforming vector, the objective of the QoS constrained formulation is to minimize the transmit power subject to SNR constraints for the individual users. Moreover, the authors of \cite{sidiropoulos2006transmit} show that the solutions to both problems are equivalent up to a scaling factor.

The more general case with multiple cochannel multicast groups is considered in \cite{karipidis2008quality}. 
Unlike the single-group case, the QoS constrained and MMF versions of the multi-group multicast beamforming problem are different in the sense that a solution to one version cannot generally be obtained by scaling a solution to the other. 
However, algorithms for the QoS constrained formulation can be straightforwardly extended to approximate the MMF version, by performing a bisection search over the target signal-to-interference plus noise ratio (SINR) values. In this paper, we will therefore restrict our attention to the QoS constained formulation.

The QoS-constrained multi-group multicast beamforming problem is a well-studied nonconvex quadratically constrained quadratic programming (QCQP) problem, for which various algorithmic approximations have been proposed. Existing approaches such as semidefinite relaxation with Gaussian randomization and successive convex approximation (SCA) algorithms --  also known as convex-concave-procedures (CCP) -- involve solving a sequence of convex subproblems.
Solutions to these subproblems can be approximated either using off-the-shelf interior-point methods or using first-order algorithms such as the alternating direction method of multipliers (ADMM). 
While the use of interior-point methods typically results in a high computational complexity, the ADMM can require a large number of iterations to achieve a certain accuracy. 
Regardless of the algorithm used to approximate each subproblem, the CCP results in nested approximation loops. 
Terminating the inner iteration after a finite number of steps can hinder the feasibiltiy of estimates, which is required to ensure that the CCP converges.
By contrast, if we assume the singular value decomposition of a matrix to be computable,\footnote{The convergence of algorithms for computing the singular value decomposition is well-studied (see, e.g., \cite{van1983matrix}).} the algorithm proposed in this paper is free of nested optimization loops.

In cases where constrained minimization becomes too costly, the superiorization methodology (see, e.g.,\cite{herman2012superiorization}, \cite{censor2015weak}) constitutes a promising alternative. 
Whereas the goal of constrained minimization is to find a feasible point (i.e., a point satisfying all constraints) for which the objective value is minimal, superiorization typically builds upon a simple fixed-point algorithm that produces a sequence of points which provably converges to a feasible point. This fixed-point algorithm serves as the so-called \emph{basic algorithm}, which is then modified by adding small perturbations in each iteration with the intent to find a feasible point with reduced (not necessarily minimal) objective value. By showing that the basic algorithm is bounded perturbation resilient, its convergence guarantee towards a feasible point can be extended to this modified algorithm called a superiorized version of the basic algorithm.

In this paper, we consider the QoS-constrained multi-group multicast beamforming problem in \cite{karipidis2008quality} with optional per-antenna power constraints as introduced in \cite{chen2017admm}. We propose an algorithmic approximation based on superiorization of a bounded perturbation resilient fixed point mapping.
To do so, we formulate the problem in a product Hilbert space composed of subspaces of Hermitian matrices. 
This allows us to approximate a feasible point of the relaxed problem with the well-known projections onto convex sets (POCS) algorithm \cite{stark1998vector}, which iteratively applies a fixed-point mapping comprised of the (relaxed) projections onto each constraint set.
We show that this operator is bounded perturbation resilient, which allows us to add small perturbations in each iteration with the intent to reduce the objective value and the distance to the nonconvex rank-one constraints.
Simulations show that, compared to existing methods, the proposed approach can provide better approximations at a lower computational cost in many cases.

\subsection{Preliminaries and Notation}\label{sec:notation}
Unless specified otherwise, lowercase letters denote scalars, lowercase letters in bold typeface denote vectors, uppercase letters in bold typeface denote matrices, and letters in calligraphic font denote sets. The sets of nonnegative integers, nonnegative real numbers, real numbers, and complex numbers are denoted by $\NN$, $\RR_+$, $\RR$, and $\CC$, respectively.
The real part, imaginary part, and complex conjugate of a complex number $x\in\CC$ are denoted by $\Re\{x\}$, $\Im\{x\}$, and $x^\ast$, respectively. The nonnegative part of a real number $x\in\RR$ is denoted by $\relu[x]\triangleq\max(x,0)$.

We denote by $\Id$ the identity operator and by $\I_N$ the $N\times N$-identity matrix. The all-zero vector or matrix is denoted by $\Null$ and the $i$th Cartesian unit vector is denoted by $\e_i$, where the dimension of the space will be clear from the context. The Euclidean norm of a real or complex column vector $\x$ is denoted by $\|\x\|_2=\sqrt{\x^H\x}$.
The $i$th singular value of a matrix $\A\in\CC^{N\times N}$ is denoted by  $\sigma_i(\A)$, where the singular values are ordered such that $\sigma_1(\A)\ge\cdots\ge\sigma_N(\A)$. 
For square matrices $\A$ we define $\diag(\A)$ to be the column vector composed of the diagonal of $\A$, and for row or column vectors $\a$ we define $\diag(\a)$ to be a square diagonal matrix having $\a$ as its diagonal. 
We write $\A\succcurlyeq \Null$ for positive semidefinite (PSD) matrices $\A$.

The distance between two points $\x,\y\in\setH$ in a real Hilbert space  $(\setH,\langle\cdot,\cdot\rangle)$ is $d(\x,\y)=\|\x-\y\|$, where $\|\cdot\|$ is the norm induced by the inner product $\langle\cdot,\cdot\rangle$. The distance between a point $\x\in\setH$ and a nonempty set $\setC\subset\setH$ is defined as $d(\x,\setC)=\inf_{\y\in\setC}\|\x-\y\|$.
Following \cite{bauschke2002phase}, we define the projection of a point $\x\in\setH$ onto a nonempty subset $\setC\subset\setH$ as the set
\begin{equation*}
\Pi_\setC(\x) = \left\{\y\in\setC|~ d(\x,\y) = d(\x,\setC)\right\},
\end{equation*}
and denote by $P_\setC:\setH\to\setC$ an arbitrary but fixed selection of $\Pi_\setC$, i.e., $(\forall \x\in\setH)$ $P_\setC(\x)\in\Pi_\setC(\x)$. If $\setC$ is nonempty, closed, and convex, the set $\Pi_\setC(\x)$ is a singleton for all $\x\in\setH$, so $\Pi_\setC$ has a unique selection $P_\setC$, which itself is called a projector. For closed nonconvex sets $\setC\neq\emptyset$ in finite-dimensional Hilbert spaces, $\Pi_\setC(\x)$ is nonempty for all $\x\in\setH$, although it is not generally a singleton. Nevertheless, we will refer to the selection $P_\setC$ as the projector, as the distinction from the set-valued operator $\Pi_\setC$ will always be clear.

A fixed point of a mapping $T:\setH\to\setH$ is a point $\x\in\setH$ satisfying $T(\x)=\x$. The set $\fix(T)=\{\x\in\setH~|~ T(\x)=\x\}$ is called the fixed point set of $T$ \cite{yamada2005adaptive}. 
Given two mappings $T_1, T_2:\setH\to\setH$, we use the shorthand $T_1T_2:=T_1\circ T_2$ to denote their concatenation, which is defined by the composition $(\forall\x\in\setH)$ $T_1T_2(\x):=(T_1\circ T_2)(\x) = T_1\left(T_2(\x)\right)$. 

For the following statements, let $\left(\setH,\langle\cdot,\cdot\rangle\right)$ be a real Hilbert space with induced norm $\|\cdot\|$. 
\begin{definition}
	A mapping $T:\setH\to\setH$ is called \emph{nonexpansive} if $(\forall\x,\y\in\setH)$ $\|T(\x)-T(\y)\|\le \|\x-\y\|$ \cite{yamada2005adaptive}.
\end{definition}
\begin{definition}\label{def:alpha_avg_nonexpansive}
	A mapping $T: \setH\rightarrow \setH$ is $\alpha$-averaged nonexpansive if there exist $\alpha\in(0,1)$ and a nonexpansive operator $R: \setH\rightarrow \setH$ such that $T = (1-\alpha)\Id + \alpha R$ \cite[Definition 4.33]{bauschke2011convex}.
\end{definition}
\begin{fact}\label{fact:nonexpansive_composition}
	Let  $T_1,\dots,T_L:\setH\to\setH$ be (averaged) nonexpansive mappings with at least one common fixed point. Then the composition $T_1\cdots T_L$ is also (averaged) nonexpansive and $\fix(T_1\cdots T_L)=\bigcap_{l\in\{1,\dots,L\}}\fix(T_l)$.
	\cite[Fact~1]{yamada2005adaptive}, \cite[Proposition~2.3]{he2017perturbation}
\end{fact}
\begin{fact}\label{fact:mann_iteration}
	Let $T:\setH\to\setH$ be a nonexpansive mapping with $\fix(T)\neq\emptyset$. Then for any initial point $\x_0\in\setH$ and $\alpha\in(0,1)$, the sequence $(\x_n)_{n\in\NN}\subset\setH$ generated by 
	\begin{equation*}
	\x_{n+1} = (1-\alpha)\x_n+\alpha T(\x_n)
	\end{equation*}
	converges weakly\footnote{In finite dimensional Hilbert spaces, weak convergence implies strong convergence \cite{yamada2011minimizing}.} to an unspecified point in $\fix(T)$.
	This fact is a special case of	
	\cite[Proposition~17.10b]{yamada2011minimizing}.
\end{fact}

%% file: 02-problem_statement.tex
\section{Problem Statement}
In Section~\ref{sec:system_and_original_problem}, we define the system model and state the multi-group multicast beamforming problem with QoS- and per-antenna-power-constraints, and we reformulate it in terms of a nonconvex semidefinite program (SDP).
A well-known approach to approximating solutions to such problems resorts to solving a convex relaxation: First, the original problem is relaxed and solved using, e.g., interior point methods. Subsequently, randomization techniques are applied to obtain candidate solutions to the original problem \cite{karipidis2008quality}, \cite{luo2010semidefinite}. 
However, in real-time applications, the complexity of interior point solvers becomes prohibitive as it grows very fast with the system size (i.e., the number of users and the number of antennas).

Therefore, in Section~\ref{sec:SDR_hilbert}, we formulate the problem in a real product Hilbert space composed of complex (Hermitian) matrices. This formulation makes the problem accessible by a variety of first-order algorithms with low complexity and provable convergence properties.

\subsection{System Model and Original Problem}\label{sec:system_and_original_problem}
Following the system model in \cite{karipidis2008quality}, we consider the downlink in a network with a transmitter equipped with $N$ antenna elements, each of them represented by an element of the set $\setN\triangleq\{1,\dots,N\}$. Each user $k\in\setK\triangleq \{1,\dots,K\}$ is equipped with a single receive antenna. The users are grouped into $M$ disjoint multicast groups $\Group_m\subseteq\setK$ indexed by $m\in\setM\triangleq\{1,\dots,M\}$, such that $\bigcup_{m=1}^M\Group_m=\setK$. Each member of a multicast group $\Group_m$ is intended to receive the same information-bearing symbol $x_m\in\CC$.
The receive signal for the $k$th user can be written as $y_k = \sum_{m=1}^M \w_m^H \h_k x_m + n_k$, where $\w_m\in\CC^N$ is the beamforming vector for the $m$th multicast group, $\h_k\in\CC^N$ is the instantaneous channel to user $k$, and $n_k\in\CC$ --- drawn independently from the distribution $\CN(0,\sigma_k^2)$ --- is the noise sample at the receiver. Consequently, the transmit power for group $\Group_m$ is proportional to $\|\w_m\|_2^2$.

In this paper, we consider the multi-group multicast beamforming problem with QoS-constraints \cite{karipidis2008quality}, which has the objective to minimize the total transmit power subject to constraints on the QoS expressed in terms of SINR requirements. We use the following problem formulation from \cite{chen2017admm}, with an individual power-constraint for each transmit antenna:
\begin{subequations}\label{eq:original_problem}
	\begin{alignat}{3}
	\underset{\{\w_m\in\CC^N\}_{m=1}^M}{\minimize} &\ \sum\limits_{m=1}^M\|\w_m\|_2^2\label{eq:original_prob_a}\\
	\st\quad &   (\forall m \in\setM)(\forall k \in\Group_m)\nonumber\\
	&\frac{ |\w_m^H\h_k|^2}{\sum_{l\neq m}|\w_l^H\h_k|^2+\sigma_k^2}\ge \gamma_k\label{eq:original_prob_b}\\
	&(\forall i\in\setN)\ \sum_{m=1}^M\w_m^H\E_i\w_m\le p_i\label{eq:original_prob_c}
	%	&(\forall i\in\setN)\ \sum_{m=1}^M\left|\w_m^H\e_i\right|^2\le p_i
	\end{alignat}
\end{subequations}
The objective function in \eqref{eq:original_prob_a} corresponds to the total transmit power. The inequalities in \eqref{eq:original_prob_b} constitute the SINR-constraints, where $\gamma_k$ is the SINR required by user $k$. The inequalities in \eqref{eq:original_prob_c} correspond to the per-antenna power constraints, where $\e_i\in\RR^N$ is the $i$th Cartesian unit vector.

The problem in \eqref{eq:original_problem} is a nonconvex QCQP, which is known to be NP-hard \cite{sidiropoulos2006transmit}.
A well-known strategy for approximating solutions to such problems is the semidefinite relaxation technique \cite{karipidis2008quality}, \cite{luo2010semidefinite}. By this technique, we obtain a convex relaxation of the original problem by reformulating it as a nonconvex semidefinite program and by dropping the nonconvex rank constraints.
More precisely, using the trace identity $\trace(\A\B)=\trace(\B\A)$ for matrices $\A,\B$ of compatible dimensions, we can write $\|\w_m\|_2^2=\w_m^H\w_m=\trace(\w_m^H\w_m)=\trace(\w_m\w_m^H)$ and $|\w_m^H\h_k|^2=\w_m^H\h_k(\w_m^H\h_k)^*=\trace(\w_m^H\h_k\h_k^H\w_m)=\trace(\w_m\w_m^H\h_k\h_k^H)$. By defining $(\forall k\in\setK)$ $\Q_k=\h_k\h_k^H$, and replacing the expression $\w_m\w_m^H$ by a positive semidefinite rank-one matrix $\X_m\in\CC^{N\times N}$ for all $m\in\setM$, we obtain the nonconvex semidefinite program
\begin{subequations}\label{eq:SDP}
	\begin{alignat}{5}
\underset{\{\X_m\in\CC^{N\times N}\}_{m=1}^M}{\minimize}\ & \sum\limits_{m=1}^M\trace(\X_m)\label{eq:SDPa}\\
\st\quad &   (\forall m \in\setM) (\forall k \in\Group_m)\label{eq:SDPb}\\
& \trace(\Q_k\X_m)\ge \gamma_k\sum\limits_{l\neq m}\trace(\Q_k\X_l) + \gamma_k\sigma_k^2\notag\\
&(\forall i\in\setN)\ \sum_{m=1}^M\trace(\E_i\X_m)\le p_i\label{eq:SDPc}\\
&(\forall m\in\setM)\ \X_m\succcurlyeq\Null\label{eq:SDPd}\\
&\rank(\X_m)\le 1,\label{eq:SDPe}
\end{alignat}
\end{subequations}
This formulation is equivalent to \eqref{eq:original_problem} in the sense that $\{\X_m=\w_m\w_m^H\}_{m=1}^M$ solves \eqref{eq:SDP} if and only if $\{\w_m\}_{m=1}^M$ solves \eqref{eq:original_problem}.

A convex relaxation of Problem~\eqref{eq:SDP} can be obtained by simply dropping the rank-constraints in \eqref{eq:SDPe}. 
The approach in \cite{sidiropoulos2006transmit}, \cite{karipidis2008quality} solves this relaxed problem and, subsequently, generates candidate approximations for Problem~\eqref{eq:SDP} (and hence \eqref{eq:original_problem}) using randomization techniques. A solution to the relaxed problem is typically found using general-purpose interior point solvers, which results in high computational cost for large-scale problems.  In the multi-group setting \cite{karipidis2008quality}, each randomization step involves solving an additional power control problem, which further increases the computational burden.

\subsection{Problem Formulation in a Real Hilbert Space}\label{sec:SDR_hilbert}
The objective of this section is to show that Problem~\eqref{eq:SDP} can be formulated in a real Hilbert space, which enables us to approach the problem by means of efficient projection-based methods. 
To this end, we consider the $\emph{real}$ vector space $\setV\triangleq \CC^{N\times N}$ of complex $N\times N$-matrices. More precisely, we define vector addition in the usual way, 
and we restrict scalar multiplication to real scalars  $a\in\RR$, where each coefficient of a vector $\X\in\setV$ is multiplied by $a$ to obtain the vector $a\X\in\setV$. In this way, $\setV$ is a real vector space, i.e., a vector space over the field $\RR$.

If we equip the space $\setV$ with a real inner product\footnote{A proof that this function is in fact a real inner product can be found in Remark~\ref{rem:real_inner_product} in the Appendix.}
\begin{equation}\label{eq:innerProduct}
(\forall \X,\Y\in\setV)\quad 
\left\langle\X,\Y\right\rangle\triangleq\Re\left\{\trace\left(\X^H\Y\right)\right\},
\end{equation}
which induces the standard Frobenius norm 
\begin{equation*}
||\X||=\sqrt{\left\langle\X,\X\right\rangle} = \sqrt{\trace\left(\X^H\X\right)},
\end{equation*}
we obtain a \emph{real} Hilbert space $\left(\setV,\langle\cdot,\cdot\rangle\right)$.  

In the remainder of this paper, we restrict our attention to the subspace $\hilbert\triangleq \{\X\in\setV~|~ \X=\X^H\}$ of Hermitian matrices.
Following the notation in \cite{stark1998vector}, we define a product space $\Hilbert$ as the $M$-fold Cartesian product
\begin{equation*}
\Hilbert \triangleq\underset{M\text{ times}}{\underbrace{\hilbert\times\dots\times\hilbert}}
\end{equation*}
 of $\hilbert$. 
 In this vector space, the sum of two vectors $\X=\left(\X_1,\dots,\X_M\right)$ and $\Y=\left(\Y_1,\dots,\Y_M\right)\in\Hilbert$ is given by $\X+\Y :=  \left(\X_1 + \Y_1,\dots,\X_M +\Y_M\right)$ and scalar multiplication is restricted to real scalars $a\in\RR$, where $a\left(\X_1,\dots,\X_M\right):=\left(a\X_1,\dots,a\X_M\right)$.
 We equip the space $\Hilbert$ with the inner product
\begin{equation}\label{eq:InnerProduct}
\Langle \X,\Y \Rangle \triangleq\sum\limits_{m=1}^M \langle \X_m,\Y_m\rangle,
\end{equation}
which induces the norm
\begin{equation*}
\Norm[\X]^2=\Langle\X,\X\Rangle=\sum\limits_{m=1}^M \|\X_m\|^2,
\end{equation*}
where $(\forall m\in\setM)$ $\X_m\in\hilbert$ and $\Y_m\in\hilbert$.
 Consequently, $\left(\Hilbert, \Langle\cdot,\cdot\Rangle\right)$ is also a real Hilbert space.

In order to pose Problem~\eqref{eq:SDP} in this Hilbert space, we express the objective function in \eqref{eq:SDPa} and the constraints in \eqref{eq:SDPb}--\eqref{eq:SDPe} in terms of a convex function and closed sets in $\left(\Hilbert, \Langle\cdot,\cdot\Rangle\right)$ as shown below:

\begin{enumerate}
	\item The objective function in \eqref{eq:SDPa} can be written as the following
	inner product:
	\begin{equation}\label{eq:hilber_objective}
	\sum\limits_{m=1}^M\trace(\X_m) = \Langle\J,\X\Rangle,
	\end{equation}
	where $\J=(\I_N,\dots,\I_N)$. This follows from \eqref{eq:innerProduct}, \eqref{eq:InnerProduct}, and the fact that $(\forall\W\in\hilbert)$ $\Im\{\trace(\W)\}=0$.
	
	\item 	The SINR constraint for user $k\in\setK$ in \eqref{eq:SDPb} corresponds to the closed half-space
	\begin{equation}\label{eq:SINR_sets}
	\Cqk=\left\{\left.\X\in\Hilbert\right|\ \Langle\X,\Z^k\Rangle\ge \sigma_k^2 \right\},
	\end{equation}
	where $(\forall k\in\setK)$ $\Z^k\in\Hilbert$ is given by
	\begin{equation*}
	\Z^k =\Big(\underset{1,\cdots,g_k-1}{\underbrace{-\Q_k,\cdots,-\Q_k}},\underset{g_k}{\underbrace{\gamma_k^{-1}\Q_k}} ,\underset{g_k+1,\cdots,M}{\underbrace{-\Q_k,\cdots,-\Q_k}}\Big).
	\end{equation*}
	Here, we introduced indices $\{g_k\}_{k\in\setK}$ that assign to each receiver $k\in\setK$ the multicast group $\Group_m$ to which it belongs (i.e., $g_k=m$, if $k\in\Group_m$).
	
	In order to verify that the set $\Cqk$ in \eqref{eq:SINR_sets} indeed represents the SINR constraint for user $k$ in \eqref{eq:SDPb}, we rearrange\footnote{In the remainder of this paper, we use the convention that $\X_m\in\hilbert$ denotes the $m$th component matrix of an $M$-tuple $\X\in\Hilbert$.}
	\begin{equation*}
	\Langle\X,\Z^k\Rangle 
	= \frac{1}{\gamma_k}\langle \X_{g_k},\Q_k\rangle - \sum\limits_{\substack{l\in\setM\\
			l\neq g_k}}\langle\X_l, \Q_k\rangle.
	\end{equation*}
	
	Using the definition of the inner product in \eqref{eq:innerProduct}, and the fact that $(\forall\W\in\hilbert)$ $\W^H=\W$ and $\Im\{\trace(\W)\}=0$, we can rewrite the constraint $\Cqk$ as
	\begin{equation*}
	\trace(\X_{g_k}\Q_k) - \gamma_k \sum\limits_{\substack{l\in\setM\\
			l\neq g_k}}\trace(\X_l\Q_k) \ge \gamma_k\sigma_k^2,
	\end{equation*}
	which corresponds to the $k$th SINR constraint in \eqref{eq:SDPb}.
	
	\item 	The per-antenna power constraints in \eqref{eq:SDPc} are expressed by the closed convex set
	\begin{equation*}
	\Cp=\left\{\X\in\Hilbert\left|~(\forall i\in\setN)~  \Langle \D^i ,\X\Rangle \le p_i \right.\right\},
	\end{equation*}
	where 
	\begin{equation}\label{eq:defDi}
	(\forall i \in\setN)\quad \D^i\triangleq(\E_i,\dots,\E_i)\in\Hilbert.
	\end{equation}
	
	This follows immediately from \eqref{eq:innerProduct} and \eqref{eq:InnerProduct}.
	
	\item 	The PSD constraints in \eqref{eq:SDPd} correspond to the closed convex cone $\Cpsd$ given by
	\begin{equation*}
	\Cpsd = \left\{\left.(\X_1,\dots,\X_M)\in\Hilbert\right|~ (\forall m\in\setM)\  \X_m\succcurlyeq\Null \right\}.
	\end{equation*}
	
	\item The rank constraints in \eqref{eq:SDPe} can be represented by the nonconvex set
	\begin{equation}\label{eq:rank_constraint}
	\setR= \left\{\X\in\Hilbert\left|~ (\forall m\in\setM)~ \rank(\X_m) \le 1\right.\right\}.
	\end{equation}
\end{enumerate}

Consequently, we can pose Problem~\eqref{eq:SDP} as 
\begin{align}\label{eq:SDP_hilbert}
\underset{\X\in\Hilbert}{\minimize}\ &\Langle\J,\X\Rangle\\\notag
\st\quad &  (\forall k \in\setK)~	\X\in\Cqk\\\notag
&\X\in\Cp,\quad
\X\in\Cpsd,\quad
\X\in\setR.
\end{align}
The problems in \eqref{eq:SDP} and \eqref{eq:SDP_hilbert} are equivalent in the sense that $\{\X_m\in\setV\}_{m\in\setM}$ solves Problem~\eqref{eq:SDP} if and only if $(\X_1,\dots,\X_M)\in\Hilbert$ solves Problem~\eqref{eq:SDP_hilbert}.
The advantage of the formulation in \eqref{eq:SDP_hilbert} is that it enables us
to (i) streamline notation, (ii) express the updates of the algorithm proposed later in Section~\ref{sec:algorithmic_solution} in terms of
well-known projections, and (iii) simplify proofs by using results in
operator theory in Hilbert spaces, as we show in the following.

It is worth noting that all constraint sets described above are closed, so a projection onto each of the sets exists for any point $\X\in\Hilbert$. This property is crucial to derive projection-based
algorithms, such as the proposed algorithm.
In particular, note that we cannot replace the inequality in \eqref{eq:SDPe} with an equality, as commonly done in the
literature. The reason is that, with an equality, the corresponding set is not closed, as shown
in Remarks~\ref{rem:closed_rank} and \ref{rem:non_closed_rank}, and the practical implication is that the projection may not exist everywhere. Specifically, this happens whenever $\X=\left(\X_1,\dots,\X_M\right)$ satisfies $\X_m=\Null$ for some $m\in\setM$, which would leave the update
rule at such points undefined in projection-based methods. This is illustrated for the case $\X=\Null\in\Hilbert$ in Example~\ref{ex:undefined_projection} below.
\begin{remark}\label{rem:closed_rank}
	The rank constraint set $\setR$ in \eqref{eq:rank_constraint} is closed.
	
	\emph{Proof:}
	Let $\left(\X^{(n)}\right)_{n\in\NN}$ be a sequence of points in $\setR$ converging to a point $\X^\star=(\X^\star_1,\dots,\X^\star_M)\in\Hilbert$ and denote by $(\forall m\in\setM)$\allowbreak$(\forall n\in\NN)$
	$
	\X^{(n)}_m=\U_m^{(n)}\S_m^{(n)}(\V_m^{(n)})^H
	$
	the singular value decomposition of the $m$th component matrix of $\X^{(n)}$. 
	It follows from $\X^{(n)}\in\setR$ that $(\forall m\in\setM)$
	$\S_m^{(n)} = \diag([s_m^{(n)},0,\dots,0])$.
	Since a sequence of zeros can only converge to zero, the singular value decomposition $\X^\star_m=\U_m^\star\S_m^\star(\V_m^\star)^H$ of the $m$th component matrix of $\X^\star$ satisfies $\S_m^\star=\diag([s_m^\star,0,\dots,0])$ for some $s^\star_m\in\RR_+$. Therefore $(\forall m\in\setM)$ $\rank(\X_m^\star)\le 1$, so $\X^\star\in\setR$.
	\pushQED{\qed}
	The above shows that $\setR$ contains all its limit points, so it is closed.\qedhere
	\popQED
\end{remark}

\begin{remark}\label{rem:non_closed_rank}
By contrast,
\begin{equation*}
\setR^\prime= \left\{\X\in\Hilbert\left|~ (\forall m\in\setM)~ \rank(\X_m) = 1\right.\right\}
\end{equation*} 
is not a closed set, since for all $\X\in\setR^\prime$ and $\alpha\in(0,1)$, the sequence $\left(\alpha^n\X\right)_{n\in\NN}$ in $\setR^\prime$ converges to $\Null\notin\setR^\prime$.
\end{remark}

\begin{example}\label{ex:undefined_projection}
	The set-valued projection of $\Null\in\Hilbert$ onto the set $\setR^\prime$ in Remark~\ref{rem:non_closed_rank} is empty.
	
	\emph{Proof:}
	Suppose that $\Pi_{\setR^\prime}(\Null)\neq\emptyset$ and let $\Z\in\Pi_{\setR^\prime}(\Null)$, i.e., $\Z$ is any of the closest points of the set $\setR^\prime$ to the zero vector $\Null$. Since $\Pi_{\setR^\prime}(\Null)\subset\setR^\prime$, $(\forall m\in\setM)$ $\rank(\Z_m)=1$, i.e., $(\forall m\in\setM)$ $\sigma_1(\Z_m)>0$.
	Therefore, for any $\alpha\in(0,1)$, $\alpha\Z\in\setR^\prime$ and $d(\Null,\alpha\Z)<d(\Null,\Z)$, i.e., $\alpha \Z\in \setR^\prime$ is closer to the zero vector than $\Z \in \setR^\prime$, thus contradicting our assumption that $\Z$ is one of the closest
	\pushQED{\qed} 
	points in $\setR^\prime$ to the vector $\Null$.\qedhere\popQED
\end{example}

%% file: 03-algorithmic_solution.tex
\section{Algorithmic Solution}\label{sec:algorithmic_solution}
The main difficulty in solving \eqref{eq:SDP_hilbert} is the presence of the nonconvex rank constraint.
A well-known technique for approximating rank-constrained semidefinite programs using convex optimization methods is the semidefinite relaxation approach \cite{sidiropoulos2006transmit}, \cite{karipidis2008quality}, \cite{luo2010semidefinite}. This approach first solves \eqref{eq:SDP_hilbert} without the rank constraint, and then it applies heuristics to obtain rank-one approximations based on the solution to this relaxed problem.
Similarly, we can obtain a convex relaxation
\begin{align}\label{eq:SDR_hilbert}
\underset{\X\in\Hilbert}{\minimize}\ &\Langle\J,\X\Rangle\\\notag
\st\quad &  (\forall k \in\setK)~	\X\in\Cqk\\\notag
&\X\in\Cp,\quad
\X\in\Cpsd,
\end{align}
of Problem~\eqref{eq:SDP_hilbert} by dropping the nonconvex constraint set $\setR$.
In principle, we could solve this relaxed problem using first-order techniques for constrained convex minimization.
For instance, we could apply a projected (sub-)\allowbreak gradient method (see, e.g., \cite[Section~3.2.3]{nesterov2018lectures}), which interleaves (sub-)\allowbreak gradient steps for the objective function with projections onto the feasible set 
of Problem~\eqref{eq:SDR_hilbert}. However, computing the projection onto the intersection of all constraint sets in Problem~\eqref{eq:SDR_hilbert} typically requires an inner optimization loop because no simple expression for this projection is known.
As it was shown in \cite{censor2014projected}, superiorization can significantly reduce the computation time compared to the projected gradient method in some applications if the projection onto the feasible set is difficult to compute.

The superiorization methodology typically relies on an iterative process that solves a convex feasibility problem (i.e., that produces a sequence of points converging to a point within the intersection of all constraint sets) by repeatedly applying a computationally simple mapping. This iterative algorithm is called the \emph{Basic Algorithm}. Based on this Basic Algorithm, the superiorization methodology automatically produces a \emph{Superiorized Version of the Basic Algorithm}, by adding bounded perturbations to the iterates of the Basic Algorithm in every iteration. 
\begin{definition}
Let $\left(\Y^{(n)}\right)_{n\in\NN}$ be a bounded sequence in a real Hilbert space and let $\left(\beta^{(n)}\right)_{n\in\NN}$ be a sequence in $\RR_+$ such that $\sum_{n\in\NN}\beta^{(n)}<\infty$. Then $(\forall n\in\NN)$ $\beta^{(n)}\Y^{(n)}$ are \emph{bounded perturbations} \cite{censor2015weak}.
\end{definition}
The perturbations are typically generated based on subgradient steps for a given objective function, in a way that ensures the sequence of perturbations to be bounded. By showing that the Basic Algorithm is bounded perturbation resilient (i.e., that the resulting sequence is guaranteed to converge to a feasible point, even when bounded perturbations are added in each iteration), one can ensure that the sequence produced by the Superiorized Version of the Basic Algorithm also converges to a feasible point. In contrast to constrained minimization, superiorization does not guarantee that the objective value of the resulting approximation is minimal. However, the limit point of the superiorized algorithm typically has a lower objective value than the limit point of the unperturbed Basic Algorithm \cite{censor2015weak}.

To apply the superiorization methodology to Problem~\eqref{eq:SDP_hilbert}, we proceed as follows. 
In Section~\ref{sec:basic_algorithm}, we propose a Basic Algorithm by defining a mapping  $\Tstar:\Hilbert\to\Hilbert$. Given any point $\X^{(0)}\in\Hilbert$, this mapping generates a sequence of points
converging to a feasible point of Problem~\eqref{eq:SDR_hilbert} by
\begin{equation}\label{eq:basic_algorithm}
    (\forall n\in\NN)\quad \X^{(n+1)}=\Tstar\left(\X^{(n)}\right).
\end{equation}
In Section~\ref{sec:perturbations}, we define a sequence $\left(\beta^{(n)}\Y^{(n)}\right)_{n\in\NN}$ of bounded perturbations,  with  the  intent to reduce slightly (i) the objective value of Problem~\eqref{eq:SDP_hilbert} and (ii) the distance to the nonconvex rank constraint $\setR$ in every iteration. As we show in Proposition~\ref{prop:nonincreasing} below, the proposed perturbations can achieve both goals simultaneously. The  sequence  of  perturbations  yields  a  Superiorized  Version of the Basic Algorithm given by $\X^{(0)}\in\Hilbert$,
\begin{equation}\label{eq:superior_alg}
(\forall n\in\NN)\quad \nextit[\X] = T_\star\left(\current[\X] + \current[\beta] \current[\Y]\right).
\end{equation}
In Section~\ref{sec:boundedness}, we prove that the algorithm in \eqref{eq:superior_alg} converges to a feasible point of Problem~\eqref{eq:SDR_hilbert} by showing that the mapping $\Tstar$ is bounded perturbation resilient, and that $\left(\beta^{(n)}\Y^{(n)}\right)_{n\in\NN}$ is a sequence of bounded perturbations.
The relation between the proposed method and the superiorization methodology is discussed in detail in Section~\ref{sec:relation_to_superiorization}.
Finally, the proposed algorithm is summarized in Section~\ref{sec:algorithm_summary}.

\subsection{Feasibility-Seeking Basic Algorithm}\label{sec:basic_algorithm}
A feasible point for the relaxed SDP in \eqref{eq:SDR_hilbert} can be found by solving the convex feasibility problem
\begin{equation}\label{eq:sdrCFP}
\text{Find }\X\in\Hilbert\text{ such that } \X\in \setC_\star\triangleq\bigcap\limits_{k=1}^K \Cqk\cap \Cp\cap \Cpsd.
\end{equation}

According to Fact~\ref{fact:mann_iteration} and Definition~\ref{def:alpha_avg_nonexpansive}, given any $\X^{(0)}\in\Hilbert$, the iteration in \eqref{eq:basic_algorithm} generates a sequence of points converging to a point in $\setC_\star$
if $\Tstar$ is $\alpha$-averaged nonexpansive with $\fix(\Tstar)=\setC_\star$.
A particular case of such a mapping, which is used in the well-known projections onto convex sets (POCS) algorithm  \cite{stark1998vector}, is given by (see also Fact~\ref{fact:nonexpansive_composition})
\begin{equation}\label{eq:pocs}
\Tstar:= T^\muPsd_{\Cpsd}T^\muP_{\Cp}T^\muK_{\CqK}\dots T^\muOne_{\CqOne},
\end{equation}
where for a nonempty closed convex set $\setC\in\Hilbert$,
\begin{equation*}
\T[\setC]^\mu=\Id + \mu (\P[\setC] - \Id)
\end{equation*}
denotes the relaxed projector onto $\setC$ with relaxation parameter $\mu\in(0,2)$. The formal expressions for the projections of $\X\in\Hilbert$ onto each of the sets in \eqref{eq:sdrCFP} are given below.
\begin{enumerate}
	\item The SINR constraint sets $\Cqk\in\Hilbert$ are half-spaces, the projections onto which are given by \cite[Example~29.20]{bauschke2011convex} $(\forall k\in\setK)$\allowbreak$(\forall \X\in\Hilbert)$
	\begin{equation*}
	P_{\Cqk}(\X)=\begin{cases}
	\X,&\text{if $\X\in\Cqk$}\\
	\X + \frac{\sigma_k^2 - \Langle\X,\Z^k\Rangle}{\Norm[\Z^k]^2}\Z^k,&\text{otherwise.}
	\end{cases}
	\end{equation*}
	
	\item 	The per-antenna power constraint set $\Cp$ is an intersection of the $N$ half-spaces defined by the normal vectors $\D^i$ in \eqref{eq:defDi} for $i\in\setN$. Since these vectors are mutually orthogonal, i.e., $(\forall i\in\setN)$\allowbreak$(\forall j\in\setN\setminus \{i\})$ $\Langle\D^i,\D^j\Rangle= 0$, the projection onto $\Cp$ can be written in closed form as
	\begin{equation*}
	P_{\Cp}(\X)=
	\X + \sum_{i:p_i<\Langle\X,\D^i\Rangle} \frac{p_i - \Langle\X,\D^i\Rangle}{\Norm[\D^i]^2}\D^i.
	\end{equation*}
	This follows from \cite[Thm~4.3-1]{stark1998vector} and Halperin's Theorem (see \cite{halperin1962product}, \cite[Thm.~4.2]{ginat2018method}).
	
	\item The set $\Cpsd$ is the intersection of PSD cones in orthogonal subspaces of $\Hilbert$. The projection of $\X\in\Hilbert$ onto $\Cpsd$ is therefore given component-wise by
	\begin{equation*}
	P_\Cpsd(\X) = \left(\pPsd(\Xk[1]),\dots,\pPsd(\Xk[M])\right),
	\end{equation*}
	where, $\hilbert_+=\{\X\in\hilbert~|~ \X\succcurlyeq \Null\}$ is the cone of PSD matrices in $\hilbert$. We use the eigendecomposition $\Xk[m]=\EV_m\EW_m\EV_m^H$ with (real) eigenvalues $\EW_m=\diag(\lambda_1(\Xk[m]),\dots,\lambda_N(\Xk[m]))$ to define the projection of $\Xk[m]\in\hilbert$ onto $\hilbert_+$ as\footnote{For the case of real symmetric matrices, see, e.g.,  \cite[Lemma 2.1]{goulart2020accuracy}. The result in \cite{goulart2020accuracy} is based on \cite[Corollary~7.4.9.3]{horn2013matrix}, which assumes complex Hermitian matrices. The generalization of \cite[Lemma 2.1]{goulart2020accuracy} to complex Hermitian matrices is straightforward.}
	\begin{equation*}
	\pPsd(\Xk[m]) = \EV_m\EW^+_m\EV_m^H,
	\end{equation*}
	where $\EW^+_m\triangleq\diag\left(\relu[{\lambda_1(\Xk[m])}],\dots,\relu[{\lambda_N(\Xk[m])}]\right)$.
\end{enumerate}
According to the fundamental theorem of POCS \cite[Thm 2.5-1]{stark1998vector}, the sequence $\left(\current[\X]\right)_{n\in\NN}$ of vectors $\current[\X]\in\Hilbert$ produced by the update rule in \eqref{eq:pocs} is guaranteed to converge to a solution of the feasibility problem in \eqref{eq:sdrCFP} for any $\zeroth[\X]\in\Hilbert$, if a solution exists (i.e., if $\setC_\star\neq\emptyset$). Note that this is the case if the relaxed semidefinite program in \eqref{eq:SDR_hilbert} is feasible.
Alternatively, we can derive this convergence guarantee immediately from Remark~\ref{rem:alpha_avg} and Fact~\ref{fact:mann_iteration}.

\subsection{Proposed Perturbations}\label{sec:perturbations}
In the following, we devise perturbations that steer the iterates of the fixed point algorithm in \eqref{eq:superior_alg} towards a solution to the nonconvex problem in \eqref{eq:SDP} and \eqref{eq:SDP_hilbert}. To do so, we introduce a mapping that reduces the objective value and a mapping that reduces the distance to rank constraint sets. Then we define the proposed perturbations based on the composition of these two mappings As proven in Proposition~\ref{prop:nonincreasing} below, the resulting perturbations can achieve both goals simultaneously.

\subsubsection{Power Reduction by Bounded Perturbations}\label{sec:power_perturbations}
In the literature on superiorization, the perturbations are typically defined based on subgradient steps of the objective function (see, e.g., \cite{censor2015weak}). For the linear objective function in \eqref{eq:SDR_hilbert}, this would result in perturbations of the form
$-\alpha \left(\I_N,\dots,\I_N\right)$ 	for some $\alpha>0$. These perturbations are problematic for the problem considered here because we are interested in solutions comprised of positive semidefinite rank-one matrices, and adding these perturbations to an iterate $\X=(\X_1,\dots,\X_M)$ may result in indefinite full-rank component matrices $\X_m-\alpha\I_N$. To avoid this problem, we introduce the function $\sobjf:\Hilbert\to\RR_+$ given by
\begin{equation}\label{eq:equiv_objective}
\sobjf(\X) \triangleq \sum_{m=1}^M\|\X_m\|_\ast,
\end{equation}
where $\|\cdot\|_\ast$ is the nuclear norm. Since $\setC_\star\subset\Cpsd$ by \eqref{eq:sdrCFP}, we have $(\forall\X\in\setC_\star)(\forall m\in\setM)(\forall i\in\setN)$ $\sigma_i(\X_m)=\lambda_i(\X_m)$, where $\lambda_i(\X_m)$ and $\sigma_i(\X_m)$ denote the $i$th eigenvalue and singular value of the $m$th component matrix of $\X$, respectively. Hence we can write
\begin{align}\label{eq:surrogate_obj}
\sobjf(\X) &= \sum_{m=1}^M\sum_{i=1}^N \sigma_i(\X_m) \\\notag
&=\sum_{m=1}^M\sum_{i=1}^N \lambda_i(\X_m) = \sum_{m=1}^M\trace(\X_m).
\end{align}
Therefore, by \eqref{eq:hilber_objective}, minimizing $\sobjf$ over $\setC_\star$ is equivalent to minimizing the linear objective function in \eqref{eq:SDP_hilbert} (or \eqref{eq:SDR_hilbert}) over $\setC_\star$, in the sense that the solution sets to both formulations are the same. As we will show below, this surrogate objective function gives rise to power-reducing perturbations, which are guaranteed not to increase the rank of their arguments' component matrices (see Remark~\ref{rem:rank_reduction}).

The power-reducing perturbations are designed according to two criteria. Firstly, they should decrease the value of the surrogate function $\sobjf$. Secondly, they should not be too large in order to avoid slowing down convergence of the Basic Algorithm. 
For a given point $\X\in\Hilbert$ we derive a perturbation $\Y_\tau^\star$ satisfying these two criteria by solving the problem
\begin{equation}\label{eq:power_pert_problem}
\Y_\tau^\star:=\Y_\tau^\star(\X)\in \underset{\Y\in\Hilbert}{\argmin}~\left( \tau \sobjf(\X + \Y) + \frac{1}{2} \Norm[\Y]^2\right).
\end{equation}
Here, $\Norm[\Y]^2$ acts as a regularization on the perturbations' magnitude, and the parameter $\tau\ge0$ balances the two design criteria. The next proposition shows that $\Y_\tau^\star$  can be easily
computed.

\begin{prop}
	The unique solution to \eqref{eq:power_pert_problem} is given by
	\begin{equation}\label{eq:power_pert_opt}
	(\forall m\in\setM) \quad \Y_\tau^\star|_m = \setD_\tau(\X_m) - \X_m,
	\end{equation}
	where $\setD_\tau:\hilbert\to\hilbert$ is the singular value shrinkage operator \cite{cai2010singular}
	\begin{align}\label{eq:svd_shrink}
	\setD_\tau(\X_m)&\triangleq \U_m\setD_\tau(\SIG_m)\V_m^H,\\\notag 
	\setD_\tau(\SIG_m)&=\diag\left(\left\{\relu[{\sigma_i(\X_m)-\tau}]\right\}_{i\in\setN}\right),
	\end{align}
	and $(\forall m\in\setM)$ $\X_m=\U_m\SIG_m\V_m$ is the singular value decomposition of $\X_m$ such that $\SIG_m=\diag\left(\left\{\sigma_i(\X_m)\right\}_{i\in\setN}\right)$.
	
	\emph{Proof:}
	Denote the perturbed point for a given choice of $\tau$ by $\Z_\tau^\star:=\X+\Y_\tau^\star$. By substituting $\Y=\Z-\X$ in \eqref{eq:power_pert_problem}, we can identify this point as $\Z_\tau^\star=\prox_{\tau \sobjf}(\X)$, where the proximal mapping is given by
	\begin{equation}\label{eq:proximal_op}
	\prox_{\tau \sobjf}(\X) \in \underset{\Z\in\Hilbert}{\argmin}~ \left(\tau \sobjf(\Z) + \frac{1}{2} \Norm[\X-\Z]^2\right).
	\end{equation}
	Note that the function
	\begin{equation*}
	\tau \sobjf(\Z) + \frac{1}{2}\Norm[\X-\Z]^2 = \tau \sum_{m=1}^M \|\Z_m\|_\ast + \frac{1}{2}\sum_{m=1}^M \|\X_m-\Z_m\|^2
	\end{equation*}
	is separable over $m$. Consequently, we can compute the proximal mapping in $\eqref{eq:proximal_op}$ by solving
	\begin{equation}\label{eq:prox_subspace}
	(\forall m\in\setM)\quad \Z_\tau^\star|_m\in\underset{\Z\in\hilbert}{\argmin}~ \tau \|\Z\|_\ast + \frac{1}{2}\|\X_m-\Z\|^2.
	\end{equation}
	According to \cite[Thm. 2.1]{cai2010singular}, the unique solution to \eqref{eq:prox_subspace} is given by $\Z_\tau^\star|_m = \setD_\tau(\X_m)$.\footnote{The proof in \cite{cai2010singular} is for real matrices. However, the generalization to complex matrices is straightforward.}
	\pushQED{\qed}
	Substituting $\Y_\tau^\star=\Z_\tau^\star-\X$ yields \eqref{eq:power_pert_opt}, which is the desired result.\qedhere
\popQED
\end{prop}

By defining $(\forall\X\in\Hilbert)$
\begin{equation}
\sigma_{\max}(\X)\triangleq \max_{\substack{{m\in\setM}\\{i\in\setN}}}\sigma_i(\X_m)
\end{equation}
we can express the power-reducing perturbation for a point $\X\in\Hilbert$ as $\Y=\Tp^\alpha(\X)-\X$, where the mapping $\Tp^\alpha\triangleq\prox_{\alpha\sigma_{\max}(\X)\sobjf}$ is given component-wise by $(\forall m \in\setM)$
\begin{equation}\label{eq:op_tp}
\Tp^{\alpha}(\X)|_m =\setD_\tau(\X_m) \quad \text{with}\quad \tau= \alpha\sigma_{\max}(\X).
\end{equation} 

Note that $\Tp^0(\X) = \X$, and $(\forall\alpha\ge 1)$ $\Tp^\alpha(\X)=\Null$.
Therefore, the magnitude of the power-reducing perturbations can be controlled by choosing the parameter $\alpha\in[0,1]$. Moreover, in contrast to performing subgradient steps for the original cost function in \eqref{eq:SDP_hilbert}, applying the perturbations in \eqref{eq:op_tp} cannot increase the rank:

\begin{remark}\label{rem:rank_reduction}
	For all $\alpha\ge0$, $\Tp^\alpha$ maps any point $\X=(\X_m)_{m\in\setM}\in\Cpsd$ to a point $\Z=(\Z_m)_{m\in\setM}=\Tp^\alpha(\X)\in\Cpsd$ satisfying $(\forall m\in\setM)$ $\rank(\Z_m)\le\rank(\X_m)$.
	This follows immediately from \eqref{eq:svd_shrink}.
\end{remark}

\subsubsection{Incorporating the Rank Constraints by Bounded Perturbations}\label{sec:rank_perturbations}
Next, we define perturbations that steer the iterate towards the rank constraint set $\setR$ in \eqref{eq:rank_constraint}.
While objective functions used for superiorization are usually convex, the function $\sobjg:\Hilbert\to\RR_+$
\begin{equation}\label{eq:rank_dist}
\sobjg(\X)\triangleq d(\X,\setR),
\end{equation}
i.e., the distance to the set $\setR$, constitutes a nonconvex superiorization objective, so our approach does not follow exactly the superiorization methodology in \cite{censor2015weak} (but we can still prove convergence).

As the perturbations may steer the iterates away from the feasible set, their magnitude should not be unnecessarily large. Therefore, we choose the rank-reducing perturbations as $\Tr(\X)-\X$, where $\Tr(\X)\in\Pi_\setR(\X)$ denotes a (generalized) projection of a given point $\X\in\Hilbert$ onto the closed nonconvex set $\setR$. 
Since $\setR$ is a closed set, the set-valued projection $\Pi_\setR(\X)$ is nonempty for all $\X\in\Hilbert$. A projection onto $\setR$ can be computed by
truncating all but the largest singular value of each component matrix to zero. We formally state this fact below.

\begin{fact}
	Let $\X_m=\U_m\SIG_m\V_m^H\in\hilbert$ be the singular value decomposition of the $m$th component matrix of $\X$ with $\SIG_m=\diag(\sigma_1(\X_m),\dots,\sigma_N(\X_m))$.
	%	 and singular values $\sigma_1(\X_m)\ge\cdots\ge\sigma_N(\X_m)$.
	Then, $(\forall \X\in\Hilbert)$ the $m$th component matrix of a point $\Tr(\X)\in\Pi_\setR(\X)$ is given by \cite[Lemma~3.2]{luke2013prox}
	\begin{equation}\label{eq:proj_rank}
	\Tr(\X)|_m = \U_m\diag\left(\sigma_1(\X_m),0,\dots,0\right)\V_m^H.
	\end{equation}
\end{fact}

\subsubsection{Combining Power- and Rank Perturbations}\label{sec:combined_perturbations}
Since both $\Tp^\alpha$ in \eqref{eq:op_tp} and $\Tr$ in \eqref{eq:proj_rank} operate on the singular values of the component matrices, their composition
% $\Ts^\alpha:\Hilbert\to\Hilbert$
 is given by $(\forall m\in\setM)$
\begin{equation*}%\label{eq:full_perturbation}
\Tr\Tp^\alpha(\X)|_m=\relu[{\sigma_1(\X_m)-\alpha\sigma_{\max}(\X)}]\u_{m1} \v_{m1}^H\in\hilbert,
\end{equation*}
where, $(\forall m\in\setM)$ $\U_m=[\u_{m1},\dots,\u_{mN}]$ and $\V_m=[\v_{m1},\dots,\v_{mN}]$.
Moreover, it is easy to verify that $(\forall \X\in\Hilbert)$\allowbreak$(\forall \alpha \ge 0)$, $\Tp^\alpha\Tr(\X)=\Tr\Tp^\alpha(\X)$. 
We will now use the composition of $\Tp^\alpha$ and $\Tr$ to define a mapping $\setY_\alpha:\Hilbert\to\Hilbert$ by $\setY_\alpha:=\Tr\Tp^\alpha-\Id$, i.e., $(\forall \X=(\X_m)_{m\in\setM}\in\Hilbert)$\allowbreak$(\forall m\in\setM)$
\begin{equation}\label{eq:pert_mapping}
\setY_\alpha(\X)|_m= \relu[{\sigma_1(\X_m)-\alpha\sigma_{\max}(\X)}]\u_{m1} \v_{m1}^H - \X_m.
\end{equation}
Finally, we define the sequence $\left(\beta^{(n)}\Y^{(n)}\right)_{n\in\NN}$ of perturbations in \eqref{eq:superior_alg} by
\begin{equation}\label{eq:seq_of_pert2}
(\forall n\in\NN)\quad \Y^{(n)}\triangleq \setY_{\alpha^{(n)}}\left(\X^{(n)}\right),
\end{equation}
where $\left(\alpha^{(n)}\right)_{n\in\NN}$ is a  sequence in $[0,1]$ and $\left(\beta^{(n)}\right)_{n\in\NN}$ is a summable sequence in $[0,1]$. The following proposition shows that the perturbations in \eqref{eq:seq_of_pert2} can simultaneously reduce the objective value and the distance to the rank constraint set.

\begin{prop}\label{prop:nonincreasing}
	Let $\alpha\in\RR_+$ and $\lambda\in[0,1]$. Then each of the following holds for $\setY_\alpha:\Hilbert\to\Hilbert$ in \eqref{eq:pert_mapping}.
	\begin{enumerate}
		\item The perturbations cannot increase the distance to the set $\Cpsd$, i.e., $(\forall\X\in\Hilbert)$ $d(\X+\lambda\setY_\alpha(\X),\Cpsd)\le d(\X,\Cpsd)$. In particular, $(\forall\X\in\Hilbert)$ $\X\in\Cpsd\Rightarrow \X+\lambda\setY_\alpha(\X)\in\Cpsd$.
		\item If $\alpha>0$, the perturbations decrease the value of the function $\sobjf$ in \eqref{eq:surrogate_obj}: $\left(\forall \X\in\Hilbert\right)$ $\sobjf\left(\X+\lambda\setY_\alpha(\X)\right)<\sobjf(\X)$ whenever $\sobjf(\X)>0$.
		\item If $\alpha>0$ and $\X\in\Cpsd$, then the perturbations decrease the objective value of Problem~\eqref{eq:SDP_hilbert}, i.e., $\Langle\J,\X+\lambda\setY_\alpha(\X)\Rangle <\Langle\J,\X\Rangle$ whenever $\Langle\J,\X\Rangle>0$.
		\item If $\lambda > 0$, the perturbations decrease the distance to the rank constraint set $\setR$. More precisely, $\left(\forall \X\in\Hilbert\right)$ $\sobjg\left(\X+\lambda\setY_\alpha(\X)\right)<\sobjg(\X)$ whenever $\sobjg(\X)>0$.
	\end{enumerate}

	\emph{Proof:}
	\begin{enumerate}
		\pushQED{\qed} 
		\item This is an immediate consequence of \eqref{eq:pert_mapping}.		
		\item It follows from \eqref{eq:svd_shrink} and \eqref{eq:op_tp} that $(\forall \X\in\Hilbert)$\allowbreak$(\forall \alpha>0)$ $\sobjf(\X)>0\Rightarrow\sobjf(\Tp^\alpha(\X))<\sobjf(\X)$. Moreover, by \eqref{eq:proj_rank} we have that $(\forall \lambda\in[0,1])$ $\sobjf((1-\lambda)\X+\lambda\Tr(\X))\le\sobjf(\X)$. 
		This implies $\sobjf(\X+\lambda\setY_\alpha(\X))= \sobjf\left((1-\lambda)\X+\lambda\Tr\Tp^\alpha(\X)\right)\le\sobjf\left(\Tp^\alpha(\X)\right)<\sobjf(\X)$ whenever $\sobjf(\X)>0$.
		\item This result follows from 1) and 2), since $(\forall \X\in\Cpsd)$ $\Langle\J,\X\Rangle=\sobjf(\X)$ according to \eqref{eq:surrogate_obj}.
		\item Since $\setR$ is closed, we can write $\sobjg(\X)=d(\X,\setR)=\Norm[\X - P_\setR(\X)]=\sqrt{\sum_{m\in\setM}\sum_{i=2}^N\sigma_i^2(\X_m)}$.
		Therefore, it follows from \eqref{eq:svd_shrink} that $(\forall \X\in\Hilbert)$\allowbreak$(\forall \alpha\in\RR_+)$ $\sobjg(\Tp^\alpha(\X))\le\sobjg(\X)$.
		Moreover, by \eqref{eq:proj_rank}, $(\forall \lambda\in(0,1])$ $\sobjg(\X)>0$ implies that $\sobjg((1-\lambda)\X+\lambda\Tr(\X))<\sobjg(\X)$.
		This in turn implies $\sobjg(\X+\lambda\setY_\alpha(\X))= \sobjg\left((1-\lambda)\X+\lambda\Tr\Tp^\alpha(\X)\right)<\sobjg\left(\Tp^\alpha(\X)\right)\le\sobjg(\X)$ whenever $\sobjg(\X)>0$.	\qedhere\popQED	
	\end{enumerate}
\end{prop}
With the perturbations defined in \eqref{eq:seq_of_pert2}, the iteration in \eqref{eq:superior_alg} yields the update rule 
\begin{equation}\label{eq:superior_alg2}
(\forall n \in \NN)\quad \nextit[\X] = T_\star\left(\current[\X] + \current[\beta] \setY_{\alpha^{(n)}}\left(\current[\X]\right)\right)
\end{equation}
of the proposed algorithm, where $\X^{(0)} \in\Hilbert$ is arbitrary, $\left(\alpha^{(n)}\right)_{n\in\NN}$ is a sequence in $[0,1]$, and  $\left(\beta^{(n)}\right)_{n\in\NN}$ is a summable sequence in $[0,1]$.

\subsection{Convergence of the Proposed Algorithm}\label{sec:boundedness}
We will now examine the convergence of the proposed algorithm in \eqref{eq:superior_alg2}.
For this purpose, let $\left(\beta^{(n)}\right)_{n\in\NN}$ be a summable sequence in $[0,1]$, let $\left(\alpha^{(n)}\right)_{n\in\NN}$ be a sequence of nonnegative numbers, and denote by $\left(\Y^{(n)}\right)_{n\in\NN}$ the sequence of perturbations according to \eqref{eq:seq_of_pert2}. Then the sequence $\left(\X^{(n)}\right)_{n\in\NN}$ produced by the algorithm in \eqref{eq:superior_alg2} converges to a feasible point of Problem \eqref{eq:SDR_hilbert} for all $\X^{(0)}\in\Hilbert$. To show this, we prove the following facts.
\begin{enumerate}
	\item The mapping $\Tstar$ in \eqref{eq:pocs} is bounded perturbation resilient.
	\item The sequence $\left(\Y^{(n)}\right)_{n\in\NN}$ is bounded, such that $\left(\beta^{(n)}\Y^{(n)}\right)_{n\in\NN}$ is a sequence of bounded perturbations.
\end{enumerate}

\subsubsection{Bounded Perturbation Resilience of the Basic Algorithm}
%In order to , we can the algorithm in \eqref{eq:superior_alg}, can be used as a feasibility seeking basic algorithm for superiorization.
The operator $T_\star$ in \eqref{eq:pocs} is known to be $\alpha$-averaged (see, e.g., \cite[Example 17.12(a)]{yamada2011minimizing}). We include this fact here for completeness:
\begin{remark}\label{rem:alpha_avg}
	The operator $T_\star$ in \eqref{eq:pocs} is $\alpha$-averaged nonexpansive.
	
	\emph{Proof:}
	Note that, for every nonempty subset $\setC\subset\Hilbert$, the reflector $R_\setC= \Id + 2(P_\setC-\Id)$ is nonexpansive \cite[Corollary~4.18]{bauschke2011convex}. Therefore, according to Definition~\ref{def:alpha_avg_nonexpansive}, $\left(\forall \mu\in (0,2)\right)$ the relaxed projector
	\begin{equation*}
	T_\setC^\mu= \Id + \mu(P_\setC-\Id) = \Id + \frac{\mu}{2}(R_\setC - \Id)
	\end{equation*}
	is $\mu/2$-averaged. Further (see Fact~\ref{fact:nonexpansive_composition}), the composite of finitely many averaged mappings is $\alpha$-averaged for some
	\pushQED{\qed}
	$\alpha\in(0,1)$. \qedhere
	\popQED{}
\end{remark}
Consequently, the bounded perturbation resilience of $\Tstar$ follows directly from \cite[Thm.~3.1]{he2017perturbation}. We summarize this fact in the following Lemma.
\begin{lemma}\cite{he2017perturbation}\label{lem:bpr_of_pocs}
	 The algorithm in \eqref{eq:superior_alg} is guaranteed to converge to a point in the solution set $\setC_\star$ of the feasibility problem in \eqref{eq:sdrCFP} if $\setC_\star\neq\emptyset$ and $\left(\current[\beta]\current[\Y]\right)_{n\in\NN}$ is a sequence of bounded perturbations. 
	
	\emph{Proof:}
	\pushQED{\qed}
	The authors of \cite{he2017perturbation} have proved the bounded perturbation resilience of $\alpha$-averaged nonexpansive mappings with nonempty fix-point set in a real Hilbert space. 
	Consequently, this lemma follows from Remark~\ref{rem:alpha_avg} and \cite[Thm. 3.1]{he2017perturbation}.\qedhere\popQED
\end{lemma}

\subsubsection{Boundedness of the Perturbations}
It remains to show that the sequence $\left(\Y^{(n)}\right)_{n\in\NN}$ 
is bounded for all sequences $\left(\alpha^{(n)}\right)_{n\in\NN}$ of nonnegative numbers and $\left(\beta^{(n)}\right)_{n\in\NN}$ in $[0,1]$ such that $\sum_{n\in\NN}\beta^{(n)}<\infty$, regardless of the choice of $\X^{(0)}\in\Hilbert$.

To this end, we note that $(\forall n \in \NN)$ $\Norm[\Y^{(n)}]\le\Norm[\X^{(n)}]$ for any sequence $\left(\alpha^{(n)}\right)_{n\in\NN}$ of nonnegative numbers:
\begin{lemma}\label{lem:decreasing_norm}
	The mapping $\setY_\alpha$ in \eqref{eq:pert_mapping} satisfies 
	\begin{equation}\label{eq:rem2}
		\left(\forall\X\in\Hilbert\right)\left(\forall\alpha\in\RR_+\right)\quad \Norm[\setY_\alpha(\X)]^2\le\Norm[\X]^2.
	\end{equation}
	
	\emph{Proof:}
	Let $(\forall m\in\setM)$ $\X_m=\U_m\diag(\{\sigma_i(\X_m)\}_{i\in\setN})\V_m^H$ denote the singular value decomposition of the $m$th component matrix of $\X$.
	According to \eqref{eq:pert_mapping}, the $m$th component matrix of $\setY_\alpha(\X)$ is given by $\setY_\alpha(\X)|_m=-\U_m\S_m\V_m^H$, where $(\forall m\in\setM)$
	\begin{equation*}
	\S_m = \diag\left(\min(\sigma_1(\X_m),\tau),\sigma_2(\X_m),\dots,\sigma_N(\X_m)\right)
	\end{equation*}
	with $\tau= \alpha\sigma_{\max}(\X)$.
	Since $(\forall \W\in\hilbert)$ $\|\W\|^2=\sum_{i\in\setN}\sigma_i^2(\W)$, we can write
	\begin{align*}
	\Norm[\setY_\alpha(\X)]^2=  \sum_{m=1}^M\|\S_m\|^2\le  \sum_{m=1}^M\|\X_m\|^2=\Norm[\X]^2,
	\end{align*}
	which concludes the proof. \qed
\end{lemma}
The following known result, which is a special case of \cite[Lemma~5.31]{bauschke2011convex}, will be used in Lemma~\ref{lem:boundedness} to prove that the proposed perturbations are bounded:
\begin{fact}\label{fact:quasi_fejer}	
	Let $\left(a^{(n)}\right)_{n\in\NN}$, $\left(\beta^{(n)}\right)_{n\in\NN}$, and $\left(\gamma^{(n)}\right)_{n\in\NN}$ be sequences in $\RR_+$ such that $\sum_{n\in\NN}\beta^{(n)}<\infty$, $\sum_{n\in\NN}\gamma^{(n)}<\infty$ and
	$
	(\forall n\in\NN)\quad 
	a^{(n+1)} \le (1+\beta^{(n)})a^{(n)} + \gamma^{(n)}.
	$
	Then the sequence $\left(a^{(n)}\right)_{n\in\NN}$ converges.
\end{fact}
\begin{lemma}\label{lem:boundedness}
	Suppose that $\left(\beta^{(n)}\right)_{n\in\NN}$ is a summable sequence in $[0,1]$ and that $(\forall n\in\NN)$ $\alpha^{(n)}\ge0$. Then the sequence of perturbations $\left(\beta^{(n)}\Y^{(n)}\right)$ with $\Y^{(n)}$ defined by \eqref{eq:seq_of_pert2} is bounded.
	
	\emph{Proof:}
			We need to show that $(\exists \bound\in\RR)(\forall n\in\NN)\ \Norm[\Y^{(n)}]\le \bound$. To this end, observe that $\left(\forall \X^{(n)}\in \Hilbert\right)\left(\forall \Z\in\fix(T_\star)\right)$ it holds that
			\begin{align*}
			\Norm[\X^{(n+1)}-\Z] &= \Norm[T_\star\left(\X^{(n)}+\beta^{(n)}\Y^{(n)}\right)-\Z]\\
				&\overset{(a)}{\le} \Norm[\X^{(n)}+\beta^{(n)}\Y^{(n)}-\Z]\\
				&\overset{(b)}{\le} \Norm[\X^{(n)}-\Z] + \beta^{(n)}\Norm[\Y^{(n)}],
			\end{align*}
			where (a) follows from the nonexpansivity of $T_\star$, and (b) is a consequence of the triangle inequality.
			By Lemma~\ref{lem:decreasing_norm}, the perturbations defined in  \eqref{eq:seq_of_pert2} satisfy $(\forall n\in\NN)$ $\Norm[\Y^{(n)}]\le\Norm[\X^{(n)}]$. Consequently, applying the triangle inequality again yields
			\begin{align*}\label{eq:bounded_proof1}
			\Norm[\X^{(n+1)}-\Z]  &\le \Norm[\X^{(n)}-\Z]\\\notag
			&\qquad + \beta^{(n)}\left(\Norm[\X^{(n)}-\Z]+\Norm[\Z]\right).
			\end{align*}			
			
			By defining $(\forall n\in\NN)$ $a^{(n)}=\Norm[\X^{(n)}-\Z]$ and $\gamma^{(n)}=\beta^{(n)}\Norm[\Z]$, we can deduce from Fact~\ref{fact:quasi_fejer} that the sequence $\left(a^{(n)}\right)_{n\in\NN}$ converges. This implies that there exists $r\in\RR$ such that $(\forall n\in\NN)$ $\Norm[\X^{(n)}-\Z]\le r$.
			
			Consequently, we have 
			\begin{align*}
			(\forall n\in\NN)\quad \Norm[\Y^{(n)}] &\overset{(a)}{\le}\Norm[\X^{(n)}]
			\overset{(b)}{\le}  \Norm[\X^{(n)}-\Z] + \Norm[\Z]\\
			&\overset{(c)}{\le}r+\Norm[\Z]=:\bound
			\end{align*}
			\pushQED{\qed}
			where (a) follows from Lemma~\ref{lem:decreasing_norm}, (b) follows from the triangle inequality, and (c) follows from Fact~\ref{fact:quasi_fejer}.\qedhere
			\popQED
\end{lemma}

Combining Lemmas~\ref{lem:bpr_of_pocs} and \ref{lem:boundedness} shows that the proposed algorithm converges to a feasible point of the relaxed semidefinite program in \eqref{eq:SDR_hilbert}. This is summarized in the following proposition.

\begin{prop}
	The sequence produced by the algorithm in \eqref{eq:superior_alg} with perturbations given by \eqref{eq:seq_of_pert2} is guaranteed to converge to a feasible point of Problem~\eqref{eq:SDR_hilbert} for all $\X^{(0)}\in\Hilbert$ if $\left(\beta^{(n)}\right)_{n\in\NN}$ is a summable sequence in $[0,1]$ and $\left(\alpha^{(n)}\right)_{n\in\NN}$ is a sequence in $\RR_+$.
	
	\emph{Proof:}
	Follows immediately from Lemma~\ref{lem:bpr_of_pocs} and Lemma~\ref{lem:boundedness}.
\end{prop} 

\subsection{Relation to the Superiorization Methodology}\label{sec:relation_to_superiorization}
The authors of \cite{censor2015weak} define superiorization as follows:
\begin{quote}
\textit{'The superiorization methodology works by taking an iterative algorithm, investigating its perturbation resilience, and then, using proactively such permitted perturbations, forcing the perturbed algorithm to do something useful in addition to what it is originally designed to do.'} 
% The original unperturbed algorithm is called the “Basic Algorithm” and the perturbed algorithm is called the “Superiorized Version of the Basic Algorithm”.
\end{quote}
Although our proposed algorithm matches this informal definition,
% of a Superiorized Version of the Basic Algorithm,
there are some slight differences to the formal definition in \cite{censor2015weak}, where the perturbations are required to be nonascending vectors for a convex superiorization objective function.
\begin{definition}[Nonascending Vectors \cite{censor2015weak}]\
Given a function $\supObj:\RR^J\to\RR$ and a point $\y\in\RR^J$, a vector $\d\in\RR^J$ is said to be nonascending for $\supObj$ at $\y$ iff $\|\d\|\le 1$ and there is a $\delta>0$ such that for all $\lambda\in[0,\delta]$ we have $\supObj(\y+\lambda\d) \le \supObj(\y)$.
\end{definition}

In our case, the goal of superiorization is two-fold, in the sense that it is expressed by two separate functions $\sobjf:\Hilbert\to\RR$ and $\sobjg:\Hilbert\to\RR$.
While the function $\sobjf$ in \eqref{eq:equiv_objective} is convex, the function $\sobjg$ in \eqref{eq:rank_dist} (i.e., the distance to nonconvex rank constraint set $\setR$ in \eqref{eq:rank_constraint}) is a nonconvex function.
Moreover, we use perturbations that are not restricted to a unit ball, and therefore they are not necessarily nonascending vectors.
However, as we have shown in Proposition~\ref{prop:nonincreasing}, the proposed perturbations simultaneously reduce the values of $\sobjf$ and $\sobjg$.
Keeping these slight distinctions in mind, we will refer to the proposed algorithm in \eqref{eq:superior_alg} as \emph{Superiorized Projections onto Convex Sets}. 	

\subsection{Summary of the Proposed Algorithm}\label{sec:algorithm_summary}
The proposed multi-group multicast beamforming algorithm is summarized in Algorithm~\ref{alg:spocs}. It is defined by the relaxation parameters $\mu_1,\dots,\mu_{K+2}$ of the operator $\Tstar$ in \eqref{eq:pocs},
a scalar $a\in(0,1)$ controlling the decay of the power-reducing perturbations, a scalar $b\in(0,1)$ controlling the decay of the sequence of perturbation scaling factors, i.e., $(\forall n\in\NN)$ $\alpha^{(n)}=a^n$ and $\beta^{(n)}=b^n$. The stopping criterion is based on a tolerance value $\epsilon>0$, and a maximum number $\nmax$ of iterations.

The arguments of the algorithm are the indices $g_1,\dots,g_K$ assigning a multicast group to each user, the channel vectors $\h_1,\dots,\h_K\in\CC^N$, SINR requirements $\gamma_1,\dots,\gamma_K$, and noise powers $\sigma_1,\dots,\sigma_K$ of all users as well as the per-antenna power constraints $p_1,\dots,p_N$.
At each step, the algorithm computes a perturbation according to \eqref{eq:pert_mapping} and applies the feasibility seeking operator $\Tstar$ in \eqref{eq:pocs}. It terminates when the relative variation of the estimate falls within the tolerance $\epsilon$, or when the maximum number $\nmax$ of iterations is reached.
Finally, the beamforming vectors $\w=\{\w_m\}_{m\in\setM}$ are computed by extracting the strongest principal component
\begin{equation}\label{eq:principal}
(\forall m \in\setM)\quad \w_m=\psi(\X_m)\triangleq\sqrt{\sigma_1(\X_m)}\u_{m1},
\end{equation}
where $(\forall m \in\setM)$ $\X_m=\U_m\SIG_m\V_m^H$, $\U_m=[\u_{m1},\cdots,\u_{mN}]$, and $\SIG_m=\diag\left(\sigma_1(\X_m),\dots,\sigma_N(\X_m)\right)$.
\begin{algorithm}[H]
\caption{Superiorized Projections onto Convex Sets}\label{alg:spocs}
\begin{algorithmic}[1]
	\State \textbf{Parameters:}~ $\{\mu_k\}_{k=1}^{K+2},~ a,b\in(0,1),~ \epsilon>0,~ \nmax\in\NN$
	\State \textbf{Input:}~ $\{g_k\}_{k\in\setK}$, $\{\h_k\}_{k\in\setK}$, $\{\gamma_k\}_{k\in\setK}$, $\{\sigma_k\}_{k\in\setK}$, $\{p_i\}_{i\in\setN}$
	\State \textbf{Output:}~ $\{\w_m\in\CC^N\}_{m\in\setM}$
	\State \textbf{Initialization:}~ Choose arbitrary $\X^{(0)}\in\Hilbert$ %\Comment{E.g., $\X^{(0)}\gets\Null$}
	\For{$n=0,\dots,\nmax-1$}	
	\State $\Y^{(n)}\gets\setY_{a^n}\left(\X^{(n)}\right)$\Comment{Eq.~\eqref{eq:pert_mapping}}
	\State $\X^{(n+1)} \gets \Tstar\left(\X^{(n)} + b^n\Y^{(n)}\right)$\Comment{Eq.~\eqref{eq:pocs}}
	\If{$\Norm[\X^{(n+1)}-\X^{(n)}]<\epsilon \Norm[\X^{(n+1)}]$}
	\State \textbf{break}
	\EndIf
	\EndFor
	\State \textbf{return} $\w=\left\{\psi\left(\X_m^{(n+1)}\right)\right\}_{m\in\setM}$\Comment{Eq.~\eqref{eq:principal}}
\end{algorithmic}
\end{algorithm}

%% file: 04-numerical_results.tex
\section{Numerical Results}
In this section, we compare Algorithm~\ref{alg:spocs} (\proposed) to several other methods from the literature. We choose identical noise levels and target SINRs for all users, i.e., $(\forall k \in\setK)$ $\sigma_k=\sigma$ and $\gamma_k=\gamma$.
For each problem instance, we generate $K$ i.i.d. Rayleigh-fading channels $(\forall k\in\setK)$ $\h_k\sim\CN(\Null,\sigma^2\I_N)$.

In the first simulation, we drop the per-antenna power constraints, i.e., we set $(\forall i \in\setN)$ $p_i=\infty$, and we consider the following algorithms:
\begin{itemize}
\item The proposed method summarized in Algorithm~\ref{alg:spocs} (\proposed)
\item Semidefinite relaxation with Gaussian randomization \cite{karipidis2008quality} (\sdrgau)
\item The successive convex approximation algorithm from \cite{mehanna2014feasible}, \cite{christopoulos2015multicast} (\fppsca{})
\item The ADMM-based convex-concave procedure from \cite{chen2017admm} (\admm)
\end{itemize} 
The \proposed{} algorithm is as described in Algorithm~\ref{alg:spocs}, with parameters $a=0.95$, $b=0.999$, $\epsilon=10^{-6}$, $\nmax=10^5$. For the QoS-constraint sets, we use relaxation parameters $(\forall k\in\setK)$ $\mu_k=1.9$, and for the per-antenna power constraint set $\setP$ and the PSD constraint $\Cpsd$, we use unrelaxed projections, i.e., $\mu_{K+2}=\mu_{K+1}=1$. We initialize the \proposed{} algorithm with $\X^{(0)}=\Null$.
The convex optimization problems in the \sdrgau{} and \fppsca{} algorithms are solved with the interior point solver SDPT3 \cite{toh1999sdpt3}. The parameters of the \admm{} algorithm are as specified in \cite{chen2017admm}.
Achieving a fair comparison between these methods is difficult because the structure of the respective algorithms is quite different. 

The \sdrgau{} algorithm begins by solving the relaxed problem in \eqref{eq:SDR_hilbert}, and, subsequently, generates random candidate beamforming vectors using the \randA{} method \cite{sidiropoulos2006transmit}, \cite{karipidis2008quality}. In the multi-group setting, where $M>1$, an additional convex optimization problem (multigroup multicast power control (MMPC), \cite{karipidis2008quality}) needs to be solved for each candidate vector. If no feasible MMPC problem is found during the \randA{} procedure, we define the output of the \sdrgau{} algorithm to be $\{\psi(\X^\star_m)\}_{m\in\setM}$, where $\X^\star\in\Hilbert$ is a solution to the relaxed SDP in \eqref{eq:SDR_hilbert}.

The \fppsca{} algorithm from \cite{mehanna2014feasible} works by solving a sequence of convex subproblems. By introducing slack variables, the feasibility of each subproblem is ensured. This obviates the need for a feasible initialization point, which is typically required to ensure convergence of CCP/SCA algorithms.

The \admm{} algorithm uses an ADMM algorithm to find a feasible starting point for the CCP. Subsequently, a similar ADMM algorithm is used to approximate each subproblem of the CCP.
Because the ADMM is a first-order method, the performance of \admm{} is heavily dependent on the stopping criteria of the inner ADMM algorithm.

By contrast, the \proposed{} algorithm does not require an initialization phase, and it works by iteratively applying a sequence of operators, which can be computed in a fixed number of steps.
Therefore, we compare the performance based on computation time. Although we exclude the time required for evaluating the performance, we note that the computation time required by each of the methods severely depends on the particular implementation.

The authors of \cite{chen2017admm} assess the performance of the considered algorithms based by comparing the transmit power achieved by the resulting beamformers. However, none of the methods considered here can guarantee feasibility of the beamforming vectors, when the algorithms are terminated after a finite number of iterations. 
Furthermore, in the multi-group case, it may not be possible to scale an arbitrary candidate beamformer $\w=\{\w_m\in\CC^N\}_{m\in\setM}$ such that it satisfies all constraints in Problem~\eqref{eq:original_problem}.
In principle, we could evaluate the performance by observing both the objective value (i.e., the transmit power of the beamformers) and a measure of constraints violation such as the normalized proximity function used in \cite{censor2012effectiveness}. However, defining this measure of constraints violation is not straightforward, as the considered methods approach the problem in different spaces. Moreover, we are interested in expressing the quality of a beamforming vector by a single value to simplify the presentation.
Therefore, we will compare the performance based on the minimal SINR achieved by the beamformer $\sqrt{\bfscale}\cdot\w$ with
\begin{equation*}
\bfscale =  \min\left(\frac{\Psdr }{\sum_{m=1}^M\w_m^H\w_m}, \min_{i\in\setN}\left( \frac{p_i}{\sum_{m=1}^M |w_{im}|}\right)\right).
\end{equation*}
The scaled vector $\sqrt{\bfscale}\cdot\w$ satisfies all power constraints, and its total power is bounded by the optimal objective value $\Psdr$ of the relaxed SDP in \eqref{eq:SDR_hilbert}. More compactly, given a candidate beamformer $\w=\{\w_m\in\CC^N\}_{m\in\setM}$ for Problem~\eqref{eq:original_problem}, we assess its performance based on the function\footnote{For the sake of simplicity, we will refer to the \emph{minimal SINR achieved by the scaled beamformer} $\sqrt{\bfscale}\cdot\w$ in \eqref{eq:sinr_min} as \emph{SINR} in the following.}
\begin{equation}\label{eq:sinr_min}
\SINRmin\left(\w\right)=
\underset{k\in\setK}{\min}~ \frac{ |\w_m^H\h_k|^2}{\sum_{l\neq m}|\w_l^H\h_k|^2+\frac{\sigma_k^2 }{\bfscale}}.
\end{equation}
Since $\Psdr$ is a lower bound on the objective value of the original problem in \eqref{eq:original_problem}, it holds $(\forall \{\w_m\in\CC^N\}_{m\in\setM})$ that $\SINRmin(\w)\le\gamma$, where equality can only be achieved, if the relaxed problem in \eqref{eq:SDR_hilbert} has a solution composed of rank-one matrices.

\subsection{Performance vs. Computation Time}
\begin{figure}[H]
	\centering
	\includegraphics[scale=\plotScale]{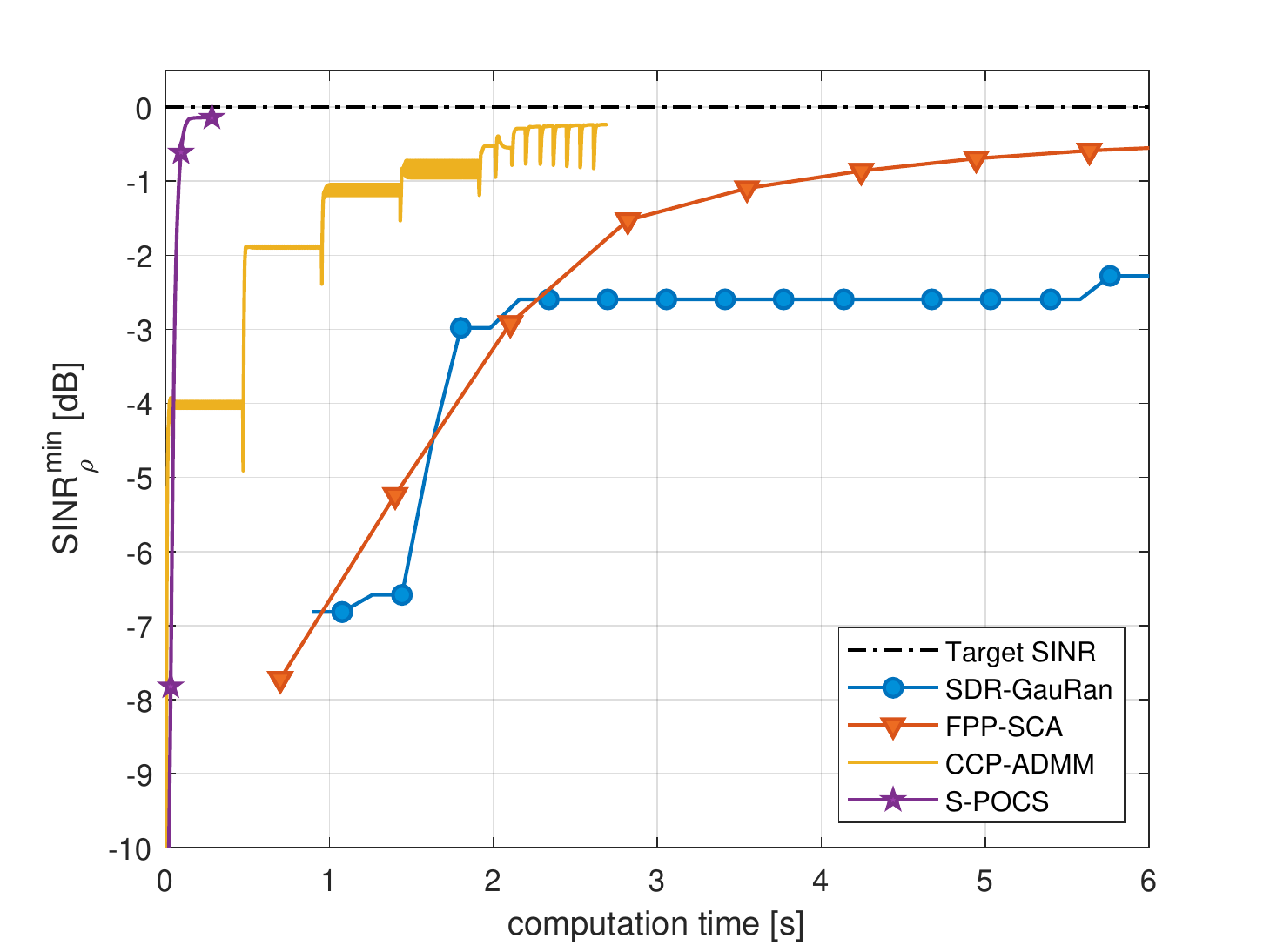}
	\caption{$\SINRmin(\w^{(t)})$ over time in a system with $N=20$ antennas and $K=20$ users users split evenly into $M=2$ multicast groups. }
	\label{fig:single_run}
\end{figure}

We will now examine how the performance metric in \eqref{eq:sinr_min} evolves over time for beamforming vectors produced by the respective algorithms.
Figure~\ref{fig:single_run} shows the performance comparison for an exemplary scenario with $N=20$ antennas, and $K=20$ users split evenly into $M=2$ groups, where $\sigma=1$, $\gamma=1$, and $(\forall i\in\setN)$ $p_i=\infty$.
It can be seen that the \proposed{} algorithm quickly converges to a point achieving an SINR close to the specified target value $\gamma$.
The discontinuities in the SINR curve for the \admm{} algorithm are due to the inner- and outer optimization loops. 
For the \sdrgau{} algorithm, the SINR increases whenever the randomization produces a beamformer with better performance than the previous one. 
The SINR of the \fppsca{} algorithm improves continuously, albeit more slowly than the \proposed{} and \admm{} algorithms.

\begin{figure}[H]
	\centering
	\includegraphics[scale=\plotScale]{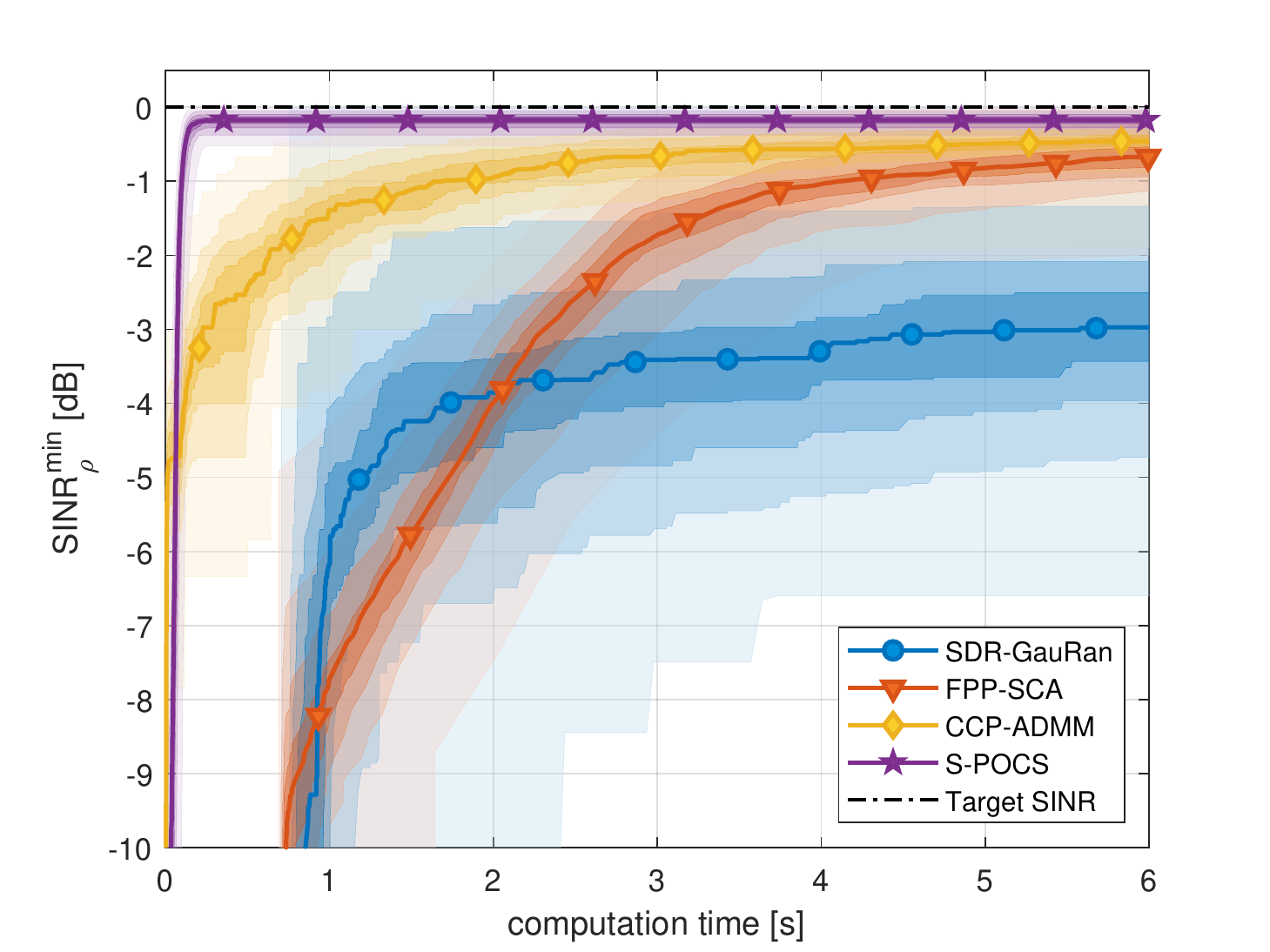}
	\caption{$\SINRmin(\w^{(t)})$ over time in a system with $N=20$ antennas and $K=20$ users split evenly into $M=2$ multicast groups. The shaded regions include the outcomes for $100\%$, $75\%$, $50\%$, and $25\%$ out of 100 problem instances, respectively,
    and the bold line represents the median.}
	\label{fig:multi_group_vs_time}
\end{figure}
Next, we evaluate the performance over 100 randomly generated problems. Since the SINR does not increase monotonically for all of the methods considered, we assume that each algorithm can keep track of the best beamformer produced so far. In this way, the oscillations in the SINR metric for the \admm{} algorithm do not have a negative impact on its average performance.

Figure \ref{fig:multi_group_vs_time} shows the performance of the beamforming vectors computed with the respective algorithms over time for a system with $N=20$ transmit antennas, and $K=20$ users split evenly into $M=2$ multicast groups.
The shaded regions correspond to the  $100\%$, $75\%$, $50\%$, and $25\%$ quantiles over all randomly generated problems. More precisely, the margins of the shaded regions correspond to the 1st, 13th, 26th, 38th, 63rd, 75th, 88th, and 100th out of 100 sorted y-axis values. For each algorithm, the median is represented by a bold line.
The \proposed{} algorithm achieves the highest median SINR, while requiring the lowest computation time among all methods considered. Moreover, it can be seen that the variation around this median value is less severe compared to the remaining approaches.
Put differently, the time required for reaching a certain SINR varies much less severely for the \proposed{} algorithm than for the remaining methods.
This can be of particular interest in delay sensitive applications, where a beamforming vector for a given channel realization must be computed within a fixed time period.

\subsection{Varying number of antennas}
In this subsection, we investigate the impact of the transmit antenna array size $N$ on the performance of the respective beamforming algorithms. To do so, we generate 100 random problem instances for each array size $N$ with $K=20$ users split evenly in to $M=2$ multicast groups. We choose unit target SINR and unit noise power for all users, and unit per-antenna power constraints, i.e., $\gamma=1$, $\sigma=1$ and $(\forall i \in\NN)$ $p_i=1$.
\begin{figure}[H]
	\centering
	\includegraphics[scale=\plotScale]{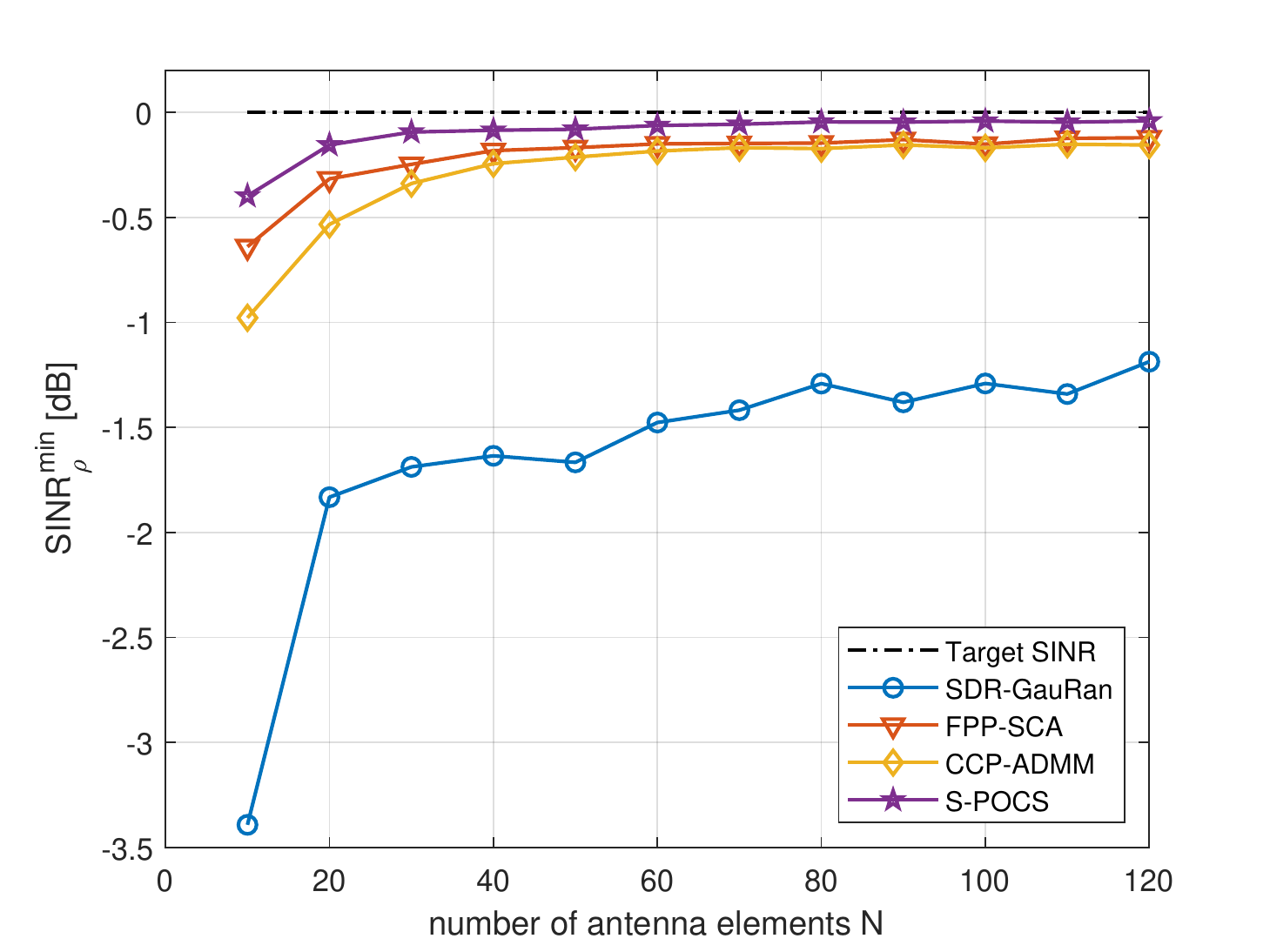}
	\caption{$\SINRmin(\w)$ for $K=20$ users split evenly into $M=2$ groups for varying antenna array sizes $N$.}
	\label{fig:multi_group_swpN}
\end{figure}
For the \sdrgau{} algorithm, we generate $200$ candidate beamforming vectors for each problem instance. We use the \admm{} algorithm with parameters as specified in \cite{chen2017admm}. Since the inner ADMM iteration converges slowly for some problem instances, we set the maximal number of steps of the ADMM to $j_{\max}=300$. For the outer CCP loop, we use the stopping criteria specified in \cite{chen2017admm}, i.e., we stop the algorithm once the relative decrease of the objective value is below $10^{-3}$ or $t_{\max}=30$ outer iterations are exceeded.
For the \fppsca{} algorithm, we use a fixed number of $30$ successive convex approximation steps.

\begin{figure}[H]
\centering
\includegraphics[scale=\plotScale]{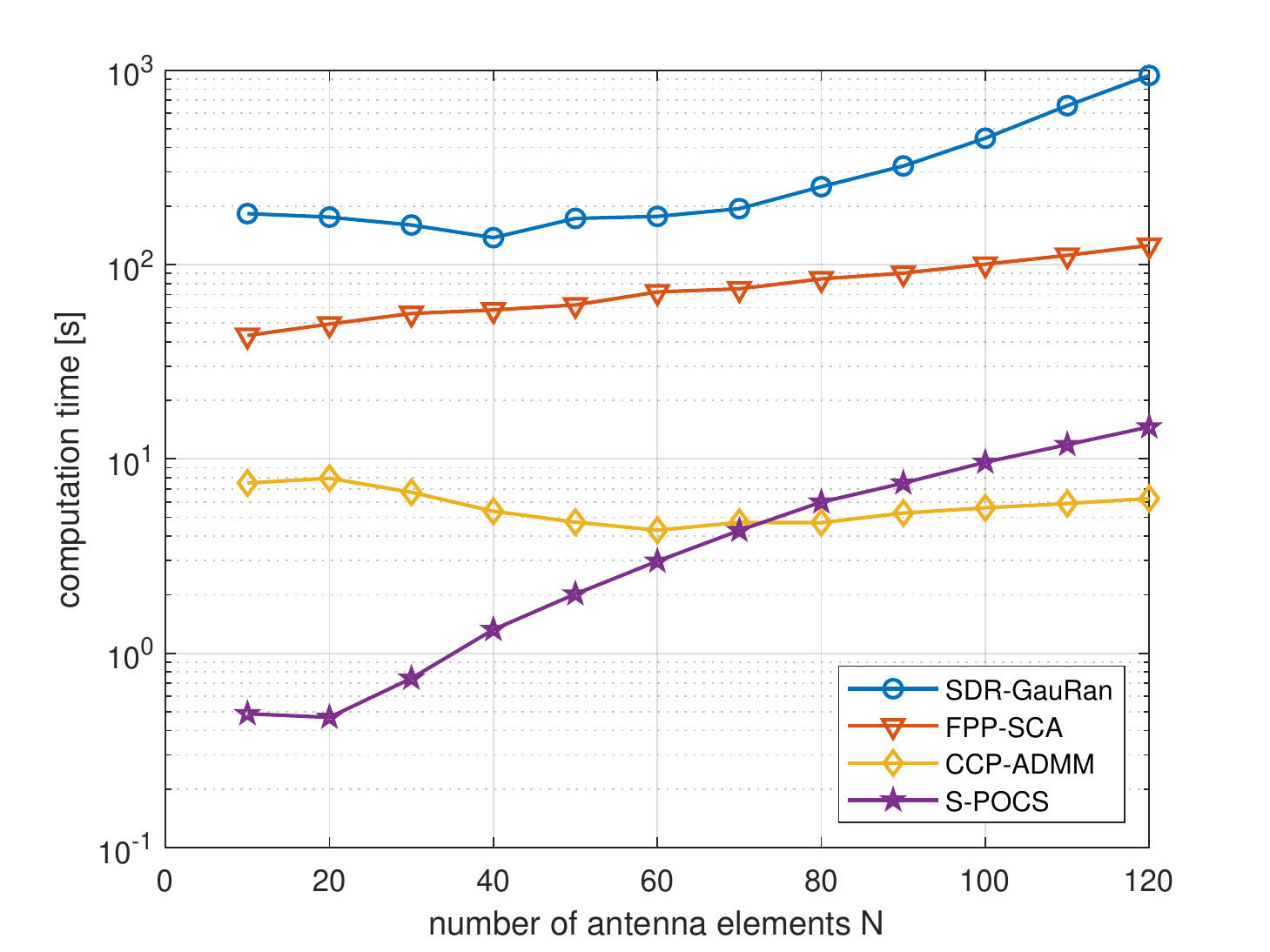}
\caption{Computation time for $K=20$ users split evenly into $M=2$ groups for varying antenna array sizes $N$.}
\label{fig:multi_group_swpN_time}
\end{figure}

Figure~\ref{fig:multi_group_swpN} shows the performance metric in \eqref{eq:sinr_min} for different numbers $N$ of transmit antennas, averaged over 100 random problem instances each. For all $N$, \proposed{} achieves highest value for $\SINRmin(\cdot)$, followed by the \fppsca{}, \admm{}, and \sdrgau{} algorithms. 
For $N\ge80$, the \proposed{} algorithm achieves an SINR of $\SINRmin(\w_{\proposed})\ge \SI{-0.05}{dB}$.
By contrast, the remaining methods do not exceed $\SINRmin(\w_{\fppsca})=\SI{-0.12}{dB}$,  $\SINRmin(\w_{\admm})\ge \SI{-0.15}{dB}$ ,$\SINRmin(\w_{\sdrgau})\ge \SI{-1.18}{dB}$, respectively.

The corresponding average computation times are shown in Figure~\ref{fig:multi_group_swpN_time}.
The \proposed{} algorithm requires
\SI{0.26}{\%}--\SI{2.38}{\%} of the computation time required by \sdrgau{},
\SI{0.95}{\%}--\SI{11.64}{\%} of the computation time required by
\fppsca{}, and
\SI{6.49}{\%}--\SI{233.6}{\%} of the computation time required by \admm{}.
For $N\ge80$, the computation time of \proposed{} exceeds that of \admm{}.

\subsection{Varying number of users}
\begin{figure}[H]
	\centering
	\includegraphics[scale=\plotScale]{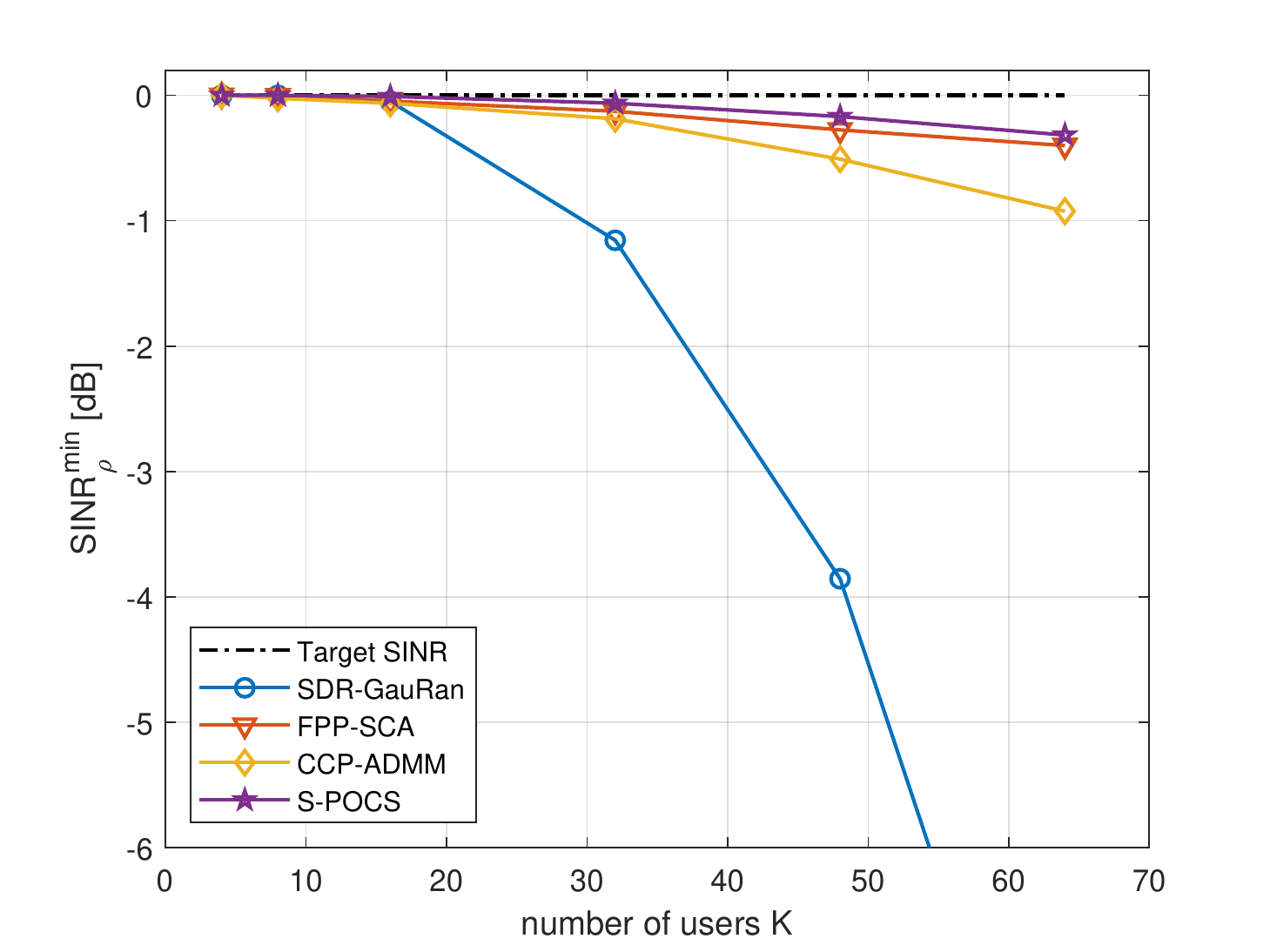}
	\caption{$\SINRmin(\w)$ for a system with $N=50$ transmit antennas and a varying number of users split evenly into $M=4$ multicast groups.}
	\label{fig:multi_group_swpK}
\end{figure}

In the following simulation, we fix an array size of $N=50$ antenna elements, and we evaluate the performance of each method for $K\in\{4, 8, 16, 32, 48, 64\}$ users split evenly into $M=4$ multicast groups. Figure~\ref{fig:multi_group_swpK} shows the performance metric in \eqref{eq:sinr_min} averaged over $100$ random problem instances for each $K$. As before, we choose $\gamma=1$, $\sigma=1$, and $(\forall i\in\setN)$ $p_i=1$.

While all algorithms achieve close to optimal performance for small numbers of users, the SINR in \eqref{eq:sinr_min} decreases considerably faster for \sdrgau{} than for the remaining methods. 
For all values of $K$, \proposed{} achieves the highest value for $\SINRmin(\cdot)$ among all methods.

\begin{figure}[H]
	\centering
	\includegraphics[scale=\plotScale]{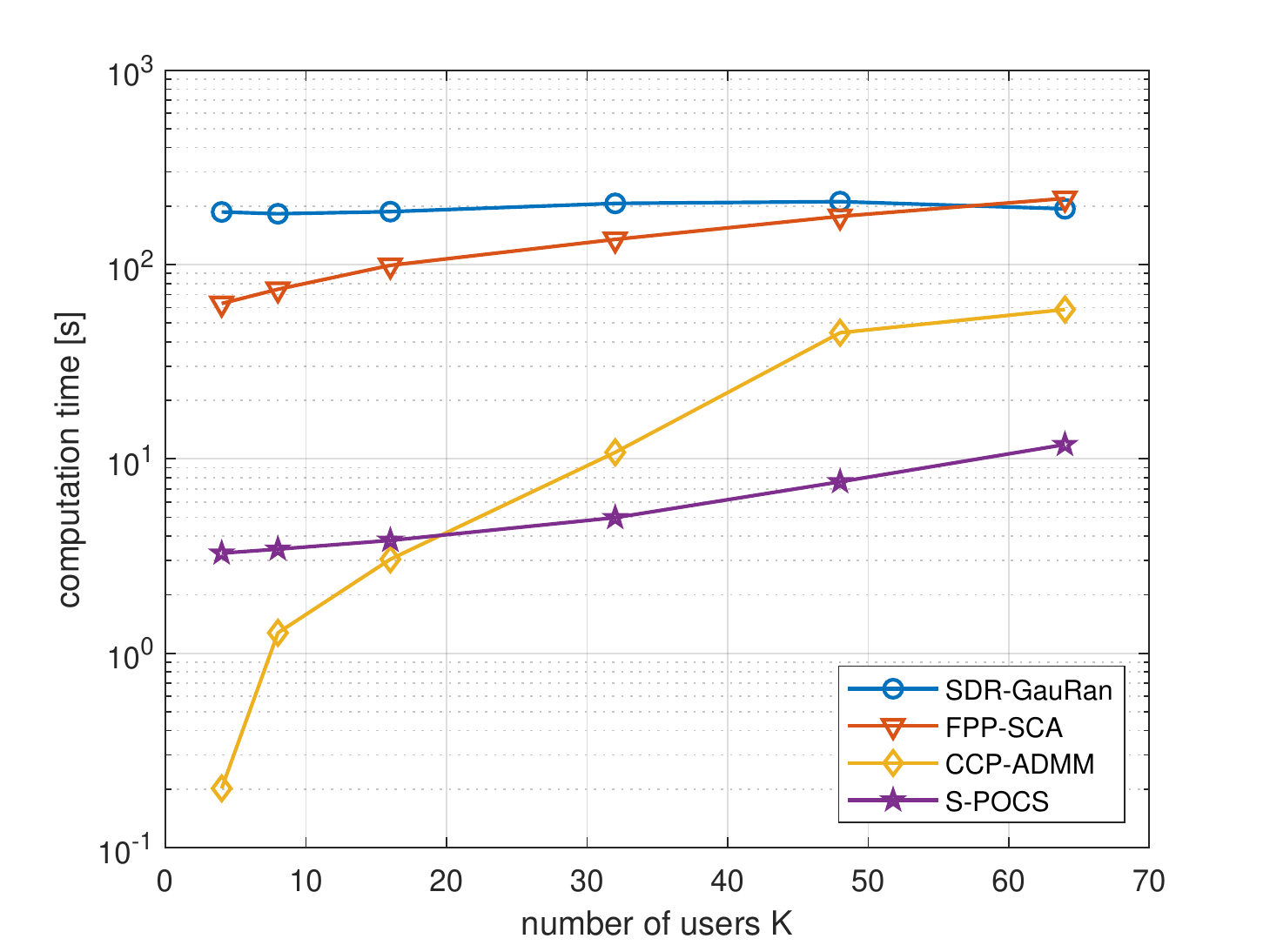}
	\caption{Computation time for a system with $N=50$ transmit antennas and a varying number of users split evenly into $M=4$ multicast groups.}
	\label{fig:multi_group_swpK_time}
\end{figure}

The corresponding average computation times are shown in Figure~\ref{fig:multi_group_swpK_time}. 
\proposed{} requires
\SI{1.76}{\%}--\SI{6.12}{\%} of the computation time required by \sdrgau{},
\SI{3.75}{\%}--\SI{5.41}{\%} of the computation time required by \fppsca{}, and
\SI{20.18}{\%}--\SI{1626}{\%} of the computation time required by \admm{}.
While the \admm{} takes only a fraction of the time required by \proposed{} for small $K$, it slows down considerably as $K$ increases. 
For moderate and large numbers of users, \proposed{} outperforms the remaining methods in terms of both approximation gap and computation time.

\subsection{Varying Target SINR}                                                     
\begin{figure}[H]
\centering
\includegraphics[scale=\plotScale]{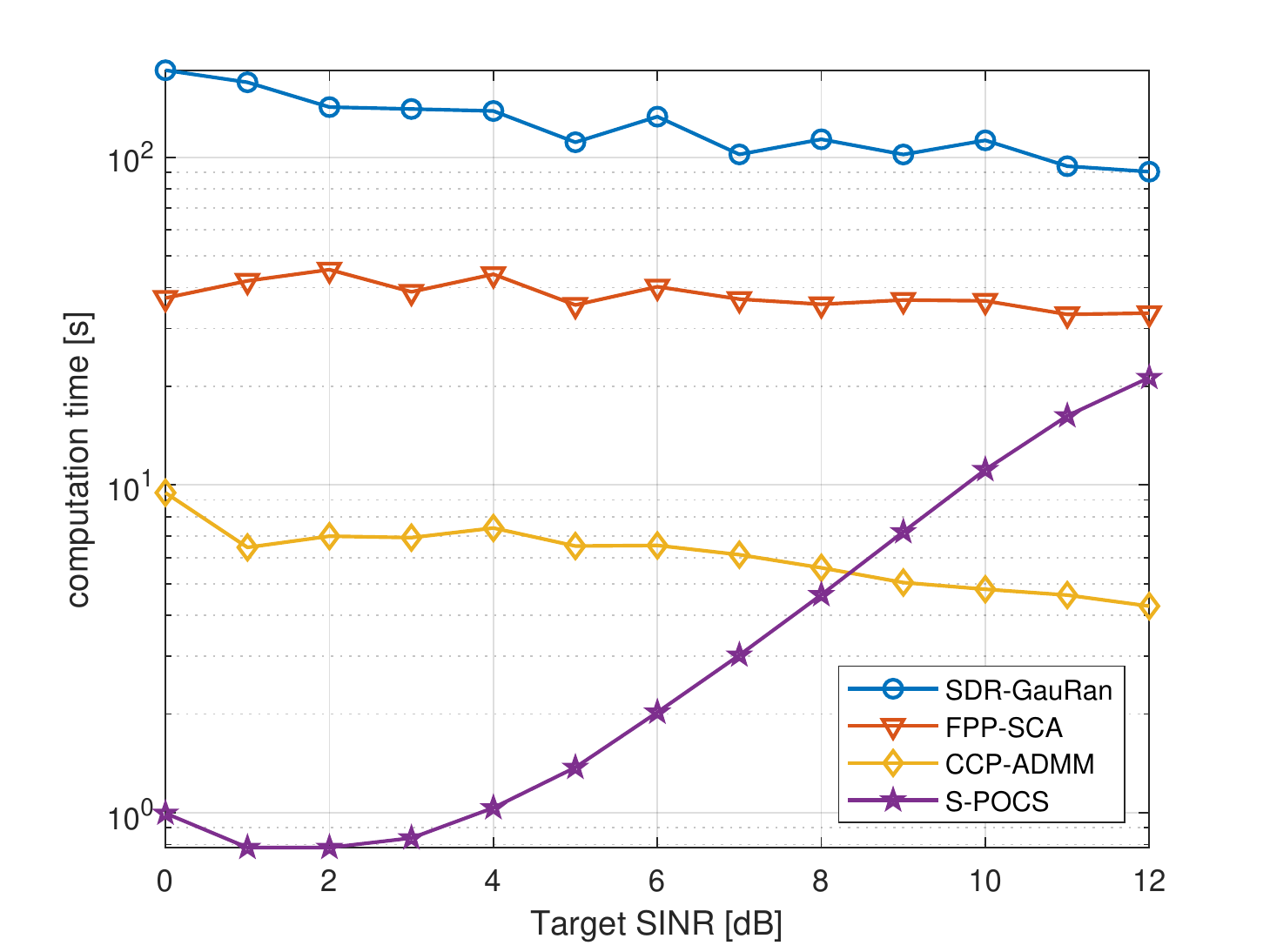}
\caption{Computation time for a system with $N=30$ transmit antennas and $K=20$ users split evenly into $M=2$ multicast groups.}
\label{fig:multi_group_swpgamma_time}
\end{figure}

In the following simulation, we evaluate the impact of the target SINR on the respective algorithms in a system with $N=30$ antenna elements, $K=20$ users split evenly into $M=2$ multicast groups, and unit noise power $\sigma=1$. Since the target SINR has a strong impact on the transmit power, we set $(\forall i\in\setN)$ $p_i=\infty$, to avoid generating infeasible instances of Problem~\eqref{eq:original_problem}.
\begin{figure}[H]
\centering
\includegraphics[scale=\plotScale]{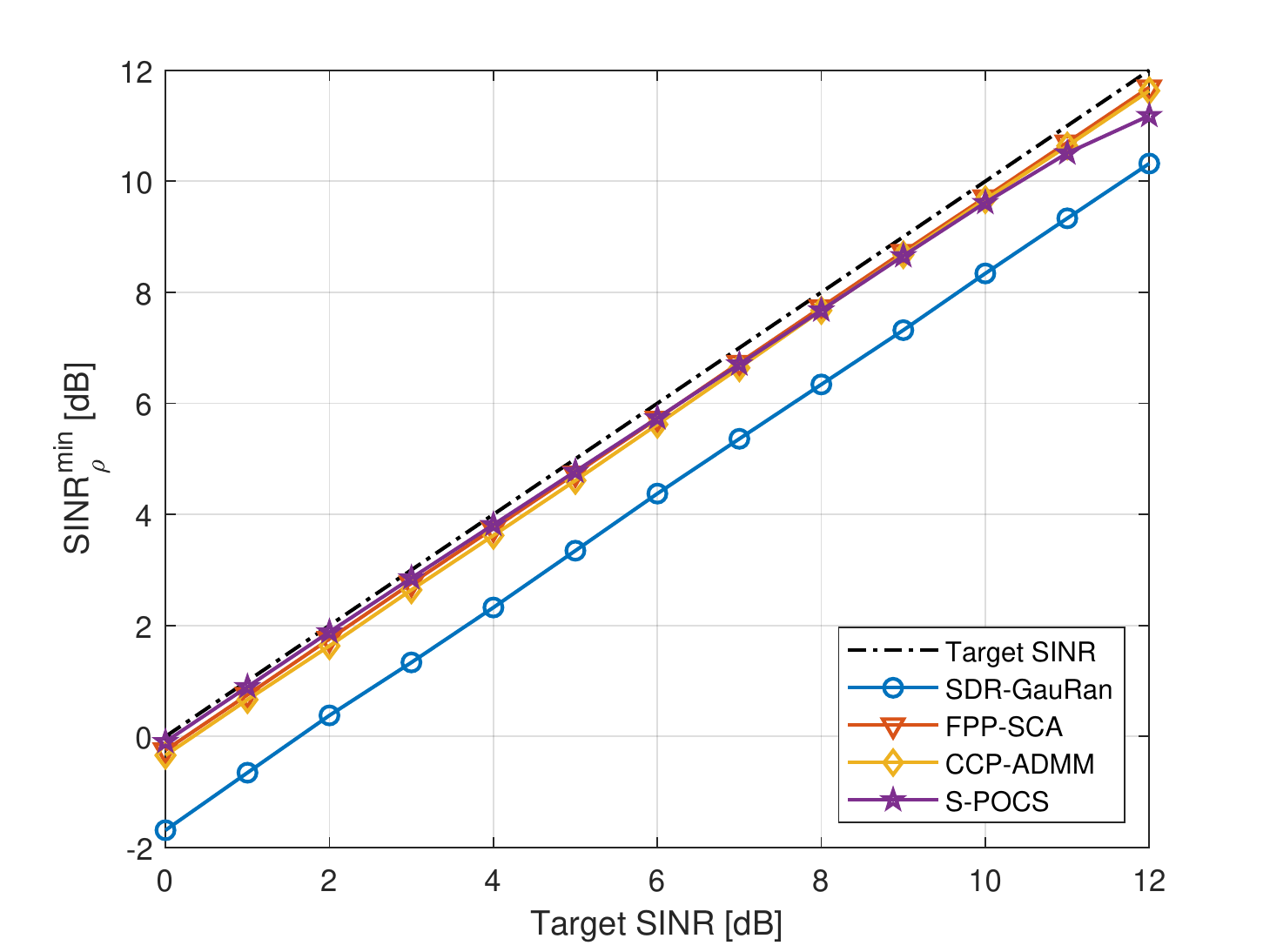}
\caption{$\SINRmin(\w)$ for a system with $N=30$ transmit antennas and $K=20$ users split evenly into $M=2$ multicast groups.}
\label{fig:multi_group_swpgamma}
\end{figure}
Figure~\ref{fig:multi_group_swpgamma} shows the performance metric in \eqref{eq:sinr_min} achieved by each method for the respective target SINR. Except for the \sdrgau{} algorithm, which exhibits a gap of about \SI{2}{dB} to the target SINR, all methods achieve close to optimal performance for each target SINR. 
Figure~\ref{fig:multi_group_swpgamma_time} shows the computation time required by each algorithm for varying target SINR $\gamma$. The average computation time of \fppsca{} is almost constant. For \sdrgau{} and \admm{}, the computation decreases slightly with an increasing target SINR. While the proposed \proposed{} algorithm converges quickly for low target SINR levels, its computation time exceeds that of the \admm{} for target SINRs above \SI{8}{dB}. This indicates that the best choice of first-order algorithms for multicast beamforming depends on the regime in which the system is operated.

%% file: 05-conclusion.tex
\section{Conclusion}
In this paper, we proposed an algorithm for multi-group multicast beamforming with per-antenna power constraints. We showed that the sequence produced by this algorithm is guaranteed to converge to a feasible point of the relaxed semidefinite program, while the perturbations added in each iteration reduce the objective value and the distance to the nonconvex rank constraints.
Numerical comparisons show that the proposed method outperforms state-of-the-art algorithms in terms of both approximation gap and computation time in many cases.
Its advantage over existing algorithms is particularly pronounced in the low target SINR regime as well as for large numbers of receivers. This makes the proposed method particularly relevant for low-energy or massive access applications. 

In comparison to other techniques, the computation time of the proposed method varies less severely across different problem instances of the same dimension. In communication systems, which are typically subject to strict latency constraints, the iteration can be terminated after a fixed number of steps without suffering severe performance loss.  Moreover, the simple structure of the proposed method allows for a straightforward implementation in real-world systems. 

The applicability of the proposed algorithm is not restricted to the multicast beamforming problem considered here. A slight modification of the rank-constraint naturally leads to an algorithm for the general rank multicast beamforming problem considered in \cite{taleb2020general}. Future research could apply superiorized projections onto convex sets to other nonconvex QCQP problems such as MIMO detection or sensor network localization \cite{luo2010semidefinite}.

%% file: 06-appendix.tex
\section{Appendix}
%%% Part 0 %%%
\begin{remark}\label{rem:real_inner_product}
	The function $\langle\cdot,\cdot\rangle$ defined in \eqref{eq:innerProduct} is a real inner product.
	
	\emph{Proof:}
	Given a real vector space $\setV$, a real inner product is a function $\langle\cdot,\cdot\rangle:\setV\times\setV\to\RR$ satisfying \cite{jain2005functional}
	\begin{enumerate}
		\item $(\forall \x\in\setV)$ $\langle\x,\x\rangle\ge0$ and $\langle\x,\x\rangle=0\iff \x=\Null$ \label{it:1}
		\item $(\forall\x,\y\in\setV)$ $\langle\x,\y\rangle= \langle\y,\x\rangle$ \label{it:2}
		\item $(\forall\x,\y\in\setV)$\allowbreak$(\forall \alpha\in\RR)$ $\langle\alpha\x,\y\rangle= \alpha\langle\x,\y\rangle$ \label{it:3}
		\item $(\forall\x,\y,\z\in\setV)$ $\langle\x+\y,\z\rangle= \langle\x,\y\rangle + \langle\y,\z\rangle$.\label{it:4}
	\end{enumerate}

Note that $(\forall\X\in\setV)$ $\Re\{\trace(\X^H\X)\}=\trace(\X^H\X)=\|\X\|_\mathrm{F}^2$, where $\|\cdot\|_\mathrm{F}$ is the standard Frobenius norm.
Consequently, \ref{it:1}) follows from the nonnegativity and positive-definiteness of a norm. The symmetry in \ref{it:2}) follows from the fact that $\trace(\A\B)=\trace(\B\A)$ for matrices $\A,\B$ with compatible dimensions, and $\Re\{\trace(\X)\}=\Re\{\trace(\X^H)\}$ for $\X\in\setV$. Moreover, \ref{it:3}) and
\ref{it:4}) follow from the linearity of $\Re\{\cdot\}$ and $\trace(\cdot)$.\qed
\end{remark}